\documentclass[prb,aps,amsmath,amssymb,showpacs,twocolumn]{revtex4-1}
\usepackage{amsmath,amssymb,graphicx,float,dcolumn,bm,color,colortbl,multirow}
\usepackage[titletoc,title]{appendix}
\usepackage[dotinlabels]{titletoc}
\usepackage{cancel}
\usepackage{braket}
\usepackage{mathrsfs}
\usepackage[mathscr]{euscript}
\usepackage{hyperref}
\usepackage{comment}
\usepackage[dvipsnames]{xcolor}

\newcolumntype{P}[1]{>{\centering\arraybackslash}p{#1}}

\newcommand{\rbkt}[1]{\left( #1\right)}
\newcommand{\sbkt}[1]{\left[ #1\right]}
\newcommand{\cbkt}[1]{\left\{ #1\right\}}

\begin{document}

\title{Real-space entanglement spectra of parton states}
\author{Abhishek Anand$^{1}$, Rushikesh A. Patil$^{1}$, Ajit C. Balram$^{2,3}$, and G. J. Sreejith$^{1}$}
\affiliation{$^{1}$Indian Institute of Science Education and Research, Pune 411008, India}
\affiliation{$^{2}$Institute of Mathematical Sciences, CIT Campus, Chennai 600113, India}
\affiliation{$^{3}$Homi Bhabha National Institute, Training School Complex, Anushaktinagar, Mumbai 400094, India}   

\begin{abstract} 
Real-space entanglement spectra (RSES) capture characteristic features of the topological order encoded in the fractional quantum Hall (FQH) states. In this work, we numerically compute, using Monte Carlo methods, the RSES and the counting of edge excitations of non-Abelian FQH states constructed using the parton theory. Efficient numerical computation of RSES of parton states is possible, thanks to their product-of-Slater-determinant structure, allowing us to compute the spectra in systems of up to 80 particles. Specifically, we compute the RSES of the parton states $\phi_2^2$, $\phi_2^3$ and $\phi_3^2$, where $\phi_n$ is the wave function of $n$ filled Landau levels (LLs), in the ground state as well as in the presence of bulk quasihole states. We then explicitly demonstrate a one-to-one correspondence of RSES of the parton states  with representations of the Kac-Moody algebras satisfied by their edge currents. We also show that for the lowest Landau level (LLL)-projected version of these parton states, the spectra matches with that obtained from the edge current algebra. We also perform a computation of spectra of the overlap matrices corresponding to the edge excitations of the parton states with a constrained number of particles in the different parton Landau levels. Counting in these match the individual branches present in RSES, providing insight about how different branches are formed.

\end{abstract}

\maketitle
\section{Introduction}
The FQH effect exhibits a rich variety of experimentally realizable phases with topological order. Nearly all the FQH phases of electrons observed in the LLL can be described in terms of integer quantum Hall (IQH) phases of emergent weakly-interacting composite fermions (CFs)~\cite{Jain89} which are electron-vortex bound states. A more general class of variational states, called the parton states~\cite{Jain89b}, has seen active interest in recent years. All of the FQH states observed in the second LL of GaAs samples can lend themselves to a parton description~\cite{Balram18, Balram18a, Balram19, Balram19a, Balram20a}. It has also been argued that parton states may be relevant to certain fractions observed in bilayer graphene~\cite{Faugno20a, Faugno20b, Balram21b}, even-denominator FQH states observed in monolayer graphene~\cite{Wu16, Kim18, Faugno20a, Faugno20b, Balram21b} and wide quantum wells~\cite{Faugno19}. Furthermore, for states in the LLL which lie beyond the realm of non-interacting CFs, candidate parton wave functions have been constructed and shown to be viable~\cite{Balram21, Balram21a, Balram21d, Dora22}. 

In the parton theory, the electron is envisaged as being made of $k$ partons, each of which goes into an IQH ground state at filling fraction $n_\beta$~\cite{Jain89b}, where $\beta$ labels the parton species. The LLL-projected $k$-parton state, labeled by a sequence of integers $(n_1,n_2,\cdots ,n_k)$, has the general form~\cite{Jain89b} 
\begin{equation}
\Psi_{\nu}^{n_1 n_2\cdots n_k} = \mathcal{P}_{\rm LLL}\prod_{\beta=1}^k \phi_{n_{\beta}}(\{z_{i}\}).
\label{eq:parton_general} 
\end{equation}
where $z_{q}=x_{q}+iy_{q}$ is the complex representation of the two-dimensional coordinate of the $q^{\rm th}$ electron, $\phi_n$ is the Slater determinant wave function of $N$ particles that completely fill the lowest $n$ LLs and $\mathcal{P}_{\rm LLL}$ projects the state into the LLL, as is appropriate in the large magnetic field limit, i.e., $B\to\infty$ . For negative $n_\beta$, the state is taken to be $\phi^*_{|n_{\beta}|}$. Since each of the partons have the same density as the parent electrons and all the partons are exposed to the external magnetic field, the $\beta^{\rm th}$ parton is associated with a charge of $(-e)\nu/n_\beta$, where $-e$ is the charge of the electron. The constraint that the charges of the partons should add up to that of electron relates the electronic filling $\nu$ to the parton filling as $\nu=\left[\sum_\beta n_\beta^{-1}\right]^{-1}$. On the spherical geometry~\cite{Haldane83}, the parton state of Eq.~\eqref{eq:parton_general} has an integral shift~\cite{Wen92} of $\mathcal{S}=\sum_\beta n_\beta$. Excitations of parton states with repeated factors of $\phi_n$, with $|n|\geq 2$ possess non-Abelian braiding statistics~\cite{Wen91}.  

The Laughlin wave function~\cite{Laughlin83} $\psi^{\rm Laughlin}_{1/(2p+1)}\equiv \phi^{2p+1}_{1}$ at filling $\nu=1/(2p+1)$ can be viewed as a parton wave function where $2p+1$ partons, each form a $n_{\beta}=1$ IQH state. The Jain CF wave functions $\psi^{\rm Jain}_{n_1/(2pn_1 \pm 1)}\equiv \mathcal{P}_{\rm LLL}\phi^{2p}_{1}\phi_{\pm n_{1}}$ at filling $\nu=n_1/(2pn_1\pm 1)$ for positive integers $n_1,\,p$ are states with $2p+1$ partons, where $2p$ of them are at filling $n_\beta=1$ and one of them is at filling $\pm n_1$. The edge theory of such a state is described by $n_1$ bosons~\cite{Wen95}. The low energy edge modes, as well as the entanglement spectrum~\cite{Li08}, show that the state counting can be interpreted as generated by $n_1$ bosonic fields. More generally, edge theory of parton states of the form $(n_1,n_2,\dots,n_k)$ which are Abelian, i.e., where none of the $n_{\beta}$'s that are greater than one repeat, are described by $c=\sum_\beta n_\beta-(k-1)$ bosons.  This can be seen by noting that each IQH state $n_{\beta}$ is made of $n_\beta$ bosonic edge modes, yielding in total $\sum_\beta n_\beta$ bosonic edge modes.  However, the density fluctuations of the $k$-partons should be identified which results in a set of $k-1$ constraints. It manifests in a reduction of the number of bosonic degrees of freedom at the edge by $k-1$, leading to above-mentioned value of chiral central charge $c$.  This procedure of gluing the unphysical partons back into the physical electrons is already implemented in the wave function of Eq.~\eqref{eq:parton_general} by identifying the coordinates of the different species of partons with the electronic coordinate, i.e., setting $z^{\beta}_{q}=z_{q}~\forall \beta$ (each parton IQH wave function is made up of \emph{all} the electrons).

This motivates us to consider the edge structure for the simplest non-Abelian parton states which contain repeating $ n_\beta$'s with $n_\beta\geq 2$.  The edge spectrum of a quantum Hall system generically depends on the confinement potential and the details of the electronic interactions. A simple approach to access the universal properties of the edge is to look at the entanglement spectrum of the bulk state~\cite{Li08}. In this work, we consider the entanglement spectra of states $(22)\equiv \phi_2^2$, $(222)\equiv \phi_2^3$ and $(33)\equiv \phi_3^2$. Unlike the states with non-repeating $n_\beta$'s where the edge theories are made of multiple bosons, the edge theories of $\phi_n^k$ are described by, edge currents that satisfy an $\widehat{su}(n)_{k}$ Kac-Moody algebra, and a $u(1)$ boson~\cite{Wen91} yielding a central charge of $k(n^2-1)/(k+n)+1$ (Eq 15.61 of Ref.~\onlinecite{DiFrancesco97}). Up to a $u(1)$ boson, the edge theory of these states is identical to that of the $k$-cluster Read-Rezayi states~\cite{Read99}. The momentum-space entanglement spectrum of the Read-Rezayi states has been previously studied in Ref.~\onlinecite{Zhu15}. We mention here that a formalism has recently been developed to rigorously study certain unprojected parton states such as $\phi_{2}^{2}$ and $\phi_{2}^{3}$~\cite{Bandyopadhyay18, Ahari22}.

The RSES of the parton wave functions has been efficiently evaluated for the case of Jain CF states, thanks to an idea originally introduced in Ref.~\onlinecite{Rodriguez12b}. The algorithm in general applies to wave functions that can be written as the product of Slater determinants or their LLL-projections, and thus can also be used to evaluate the RSES of certain parton states. In this work, we employ the algorithm to unprojected and projected partons states, as well as a few of their bulk excited states and explicitly demonstrate a one-to-one correspondence of the RSES with representations of their edge current algebra. 

The paper is organized as follows. Numerical details of the RSES computation are explained at length in Sec. \ref{sec:numericalDetails}. There we provide details of the RSES computation algorithm for parton states and discuss the prescription used for LLL-projection. In that section, we also present the calculation of edge current algebra of $\phi_n^k$ parton states using representations of $\widehat{su}(n)_k$ Kac-Moody algebra. All the results of the paper are presented in Sec. \ref{sec: results}. First, RSES for three parton states which are unprojected $\phi_2^2$, $\phi_2^3$ and $\phi_3^2$ are presented. We also present the RSES for parton states $\phi_2^2$ and $\phi_2^3$ when they have a quasihole excitation. RSES of LLL-projected $[\phi_1^2 \phi_2]^2$ is also provided. Edge spectra for unprojected $\phi_2^2$, $\phi_2^3$, and $\phi_3^2$ are presented which is computed by exact diagonalization of the overlap matrix sectors where LL occupation number is restricted. Finally, we conclude the paper with a summary of our findings in Sec. \ref{sec: conclusion}.
\section{Numerical details}
\label{sec:numericalDetails}
\subsection{RSES for parton states} \label{sec:RSESpartons}
The entanglement spectrum (ES) serves as a useful probe of the topological properties of FQH states~\cite{Li08}. It has been argued that the ES has a one-to-one correspondence with the low-energy edge spectrum providing a natural way to understand the edge theory~\cite{Li08, Bernevig2011, DRR12}. Exact analytical calculation of ES can be done for Slater determinant states including the IQH states~\cite{SierraRodriguez2009}. Direct numerical calculation of ES of correlated FQH states requires a numerical calculation of the reduced density matrix. The dimension of this matrix grows exponentially with subsystem size, making it infeasible to evaluate the spectra for large systems. 

However, for trial wave functions that can be written as a product of Slater determinants (or LLL-projections of that product), the ES across a rotationally symmetric real-space cut can be evaluated in a computationally efficient manner~\cite{RSES:simon2012, RSES:2021}. Such wave functions include the Jain CF wave functions as well the more general set of parton states; we discuss the latter in this work. 

The key computational simplification arises from the specific structure of the wave functions, which allows us to identify a small set of states whose span contains all the eigenstates of the reduced density matrix with non-zero eigenvalues. The dimension of this subspace does not scale with the subsystem size. The task then reduces to evaluating the matrix elements of the reduced density matrix between these special sets of states.
The simple structure of these basis states allows us to explicitly evaluate these matrix elements using Monte Carlo methods.
The method makes it possible to efficiently obtain the ES in a few dominant angular momentum sectors in large systems with hundreds of particles. In comparison, exact diagonalization can compute the momentum-space ES for about 20 particles~\cite{Regnault2013}. Infinite DMRG methods can produce the momentum-space ES for a nominally infinite system if the parent Hamiltonian is known. The latter is not true for the general parton states or the general Jain CF states~\cite{Sreejith18}.

The method presented here was introduced in Ref.~\onlinecite{RSES:simon2012} and has been previously employed to obtain scaling properties of the entanglement Hamiltonian for the Jain CF states~\cite{Greg2021}. Here we provide details of the method and point out the changes needed when specializing to the case of parton states.
\subsection{Construction of entanglement wave functions} 
\label{sec:EWFs}
\begin{figure}[h]
\includegraphics[width=0.4\columnwidth]{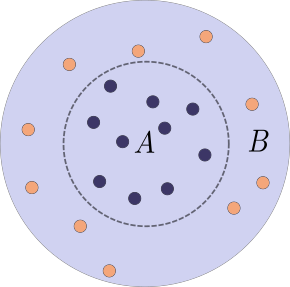}
\caption{A rotationally symmetric real-space cut on disc geometry dividing the full system into $A$ and $B$ subsystems.}
\label{fig:schematic_realspace_cut}
\end{figure}

We will work in a disc geometry with a rotationally symmetric cut, shown schematically in Fig.~\ref{fig:schematic_realspace_cut}. Since the full state has a fixed total number of particles $N$, the reduced density matrix is block diagonal in $(N_A, N_B)$ sectors, where $N_A$ and $N_B=N-N_A$ are the particles in the subsystem $A$ and $B$ on the two sides of the real-space cut.  We will work in one of these blocks where $N_{A/B}$ are close to the expected number of particles in that subsystem ($\nu/2\pi$ times the area in units of the magnetic length $\ell=\sqrt{\hbar c/(eB)}$). We assume $\ell=1$ in our calculations.
Due to rotational symmetry, the eigenstates of the reduced density matrix can be labeled by the angular momentum eigenvalues. The entanglement spectrum is obtained as the negative logarithm of the eigenvalues of the angular momentum blocks of the reduced density matrix plotted as a function of the angular momenta. 

Now we proceed to give details of the method. Any Slater determinant wave function of $N$ particles occupying angular momentum orbitals $(k_1,k_2,k_3\dots, k_N)$ can be expanded as an antisymmetrization of the product of smaller Slater determinants of $N_A$ and $N_B$ particles as follows:
\begin{multline}
S_{k_1\dots k_N}(Z)
=\\\sum_{\sigma \in P} \epsilon(\sigma)\,
 S_{k_{\sigma(1)} \dots k_{\sigma(N_A)}} (Z_A)\, S_{k_{\sigma(N_A+1)} \dots k_{\sigma(N)}} (Z_B)\label{eq:detTodetdet}
\end{multline}
where $Z_A\equiv(z_1,\dots,z_{N_A})$ and $Z_B\equiv(z_{N_A+1},\dots,z_{N})$; $P$ is the set of $(N!/N_A!N_B!)$ permutations of the ordered set $(1,\dots,N)$ such that first $N_A$ and the last $N_B$ entries of the $P(1,\dots,N)$ are in an increasing order,  and $\epsilon(\sigma)$ is the sign for a permutation represented by $\sigma$. 

Each orbital $k_i$ refers to a pair of quantum numbers $(n_i,m_i)$ namely the LL-index $n$ and angular momentum quantum-number $m$. In the disc geometry, these orbitals have the form (up to normalization):
\begin{equation}
\eta_{n,m}(z)=z^m L^m_n\rbkt{\frac{|z|^2}{2}} \text{e}^{-\frac{|z|^2}{4}}
\end{equation}
where $L^m_{n}(z)$ is the associated Laguerre polynomial.  To keep track of the signs of permutations, we will associate some order within the set of orbitals.

The product of two Slater determinants $S_M$ and $S_R$ where $M=(m_1,m_2,\dots,m_N)$ and $R=(r_1,r_2,\dots,r_N)$ 
can be written as (using Eq.~\eqref{eq:detTodetdet})
\begin{equation} \label{eq: DetToEntanglementWFs}
\psi=\sum_{\lambda} \epsilon(\lambda) \xi^{A}_\lambda (Z_A) \xi^{B}_\lambda (Z_B).
\end{equation}
Here $\lambda$ goes over all $(N!/N_A!N_B!)^2$ possible ways to split the ordered set $M$ into two disjoint ordered sets $M_A$ and $M_B$ and the ordered set $R$ into two disjoint ordered sets $R_A$ and $R_B$. Here $M_A$ and $R_A$ have sizes $N_A$ and $M_B$ and $R_B$ have sizes $N_B$. The sign $\epsilon(\lambda)$ is the sign of the permutation that takes the set $(M,R)$ into $(M_A,M_B,R_A,R_B)$. The entanglement wave functions (EWFs), $\xi$, are defined as 
\begin{align}
    \xi^{A}_\lambda &= S_{M_A}(Z_A) S_{R_A}(Z_A), \nonumber \\
    \xi^{B}_\lambda &= S_{M_B}(Z_B) S_{R_B}(Z_B).
\end{align}
The generalization to the case where the product consists of more than two Slater determinants is straightforward. The product  $S_{M^1}S_{M^2}\dots S_{M^q}$ can be written in the form of Eq.~\eqref{eq: DetToEntanglementWFs} where the EWFs are given by 
\begin{equation}
\xi^{A}_\lambda = \prod_{i=1}^q S_{M^i_A}(Z_A),\;\;\;    \xi^{B}_\lambda = \prod_{i=1}^q S_{M^i_B}(Z_B)
\end{equation}
Here $\lambda$ corresponds to one of the $(N!/N_A!N_B!)^q$ ways to split every $M^i$ into disjoint ordered subsets $M_A^i$ and $M_B^i$ of sizes $N_A$ and $N_B$.

If a wave function $\psi$ contains more than one copy of the same Slater determinant -- for instance, the unprojected Jain CF states $\phi_1^{2p}\phi_n$ -- then different splittings $\lambda$ can result in the same EWFs. 
To clarify this, consider a 10-particle state $\phi_1^2$ which contains the two copies of the Slater determinant $\phi_1=S_{(0,1,2,3\dots,9)}$ (all orbitals are in the LLL so the LL indices have been suppressed). Here $M^1=M^2=(0,1,2\dots,9)$. Two distinct splittings, given by
\begin{align}
M^1_A=(0,1,2,3,5),\;M^1_B=(4,6,7,8,9)\nonumber\\ M^2_A=(0,1,2,4,5),\; M^2_B=(3,6,7,8,9)\nonumber
\end{align}
and
\begin{align}
M^1_A=(0,1,2,4,5),\;M^1_B=(3,6,7,8,9)\nonumber\\
M^2_A=(0,1,2,3,5),\;M^2_B=(4,6,7,8,9)\nonumber 
\end{align}
result in the same EWFs $\xi_\lambda^A$ and $\xi_\lambda^B$. 
This allows simplification of Eq. \eqref{eq: DetToEntanglementWFs} to 
\begin{equation}
\psi = \sum_i s_i\, \xi_i^A (Z_A)\xi_i^B (Z_B)
\end{equation}
where $i$ indexes the distinct EWFs and $s_i$ is the sum of $\epsilon(\lambda)$ for all splittings $\lambda$ that result in the same EWFs $\xi_i^A,\, \xi_i^B$. 

\subsection{Diagonalizing $L_z^A$-blocks of $\rho_A$ } 
\label{sec:lzaBlocks}

The RSES can be calculated by diagonalizing the reduced density matrix obtained as 
$\rho_A=\sum_{j,k} s_j s_k\ket{\xi^{A}_j} \langle \xi^{B}_{k} \vert \xi^{B}_j \rangle \bra{\xi^{A}_{k}}$. Exhaustive enumeration of all splittings $\lambda$ of the resulting EWFs is computationally infeasible. However, we are typically interested only in the specific small angular momentum sectors. 
To calculate a block of reduced density matrix $\rho_A$ in a fixed $L_z^{A}$-sector, we only need to consider a restricted sum of only those EWFs which have the desired angular momentum. This is given by

\begin{equation}
\rho^{L_{z}^A}_A=\sum_{j,k}' s_j s_k\ket{\xi^{A}_j} \langle \xi^{B}_{k} \vert \xi^{B}_j \rangle \bra{\xi^{A}_{k}}
\end{equation}
where $L_z^A=\sum_i M_A^{i,\rm tot}$,  and $M_A^{i,\rm tot}$ is total momenta of $A$ subsystem for the $i^{\rm th}$ splitting. For clarity, consider the case of the $\phi_1^2$ wave function. In a given $(N_A,N_B)$-sector of $\rho_A$,  there are no EWFs with $L_z^A$ less than $L_{z,{\rm min}}^A=N_A(N_A-1)$ but  for angular momenta $0,1,2$ and $3$ above 
$L_{z, {\rm min}}^A$, the numbers of distinct EWFs are $1,2,5$ and $10$ respectively.

It can be shown that the non-zero eigenvalues of $\rho_{A}^{L_z^A}$ are the same as that of the following finite-dimensional matrix:
\begin{equation} \label{eq::defM}
    M_{jk}=\sum_{i}' s_j s_i\langle{\xi^{A}_{i}} \vert \xi^{A}_{j} \rangle \langle \xi^{B}_{i}  \vert {\xi^{B}_{k}} \rangle
\end{equation}
If the set of the EWFs are overcomplete, then $M$ has excess zero eigenvalues compared to $\rho_{A}^{L_z^A}$. The excess zero eigenvalues do not affect the entanglement spectra.

The matrix elements of $M$ can be evaluated in terms of the overlaps between the EWFs: 
\begin{align}
\langle{\xi^{A}_{i}} \vert \xi^{A}_{j} \rangle = \int_A \overline{\xi}^{A}_{i}(Z_A) \xi^{A}_{j}(Z_A) dZ_A\nonumber \\
\langle{\xi^{B}_{i}} \vert \xi^{B}_{j} \rangle = \int_B \overline{\xi}^{B}_{i}(Z_B) \xi^{B}_{j}(Z_B) dZ_B
\end{align}
Standard Metropolis Monte Carlo methods can be used to estimate these overlaps up to a proportionality constant. 
For given $N_A$, we choose one of the EWFs with smallest $L^{A}_z$ value, $\xi^{A}_0$, as the sampling wave function to estimate the overlap ratio $\langle \xi^{A}_\lambda\vert \xi^{A}_\mu \rangle/ \langle \xi^{A}_0 \vert \xi^{A}_0 \rangle$. For computing $\langle \xi^{B}_\lambda\vert \xi^{B}_\mu \rangle/ \langle \xi^{B}_0 \vert \xi^{B}_0 \rangle$, we use $\xi^{B}_0$, which is complementary EWF to  $\xi^{A}_0$. Since we use the same sampling wave function for all the $L_A$ sector, we get the overlap matrix $M$ upto a multiplicative factor of $\vert\xi^{A}_0 \xi^{A}_0 \vert^{-2}$. This also implies that the entire entanglement spectrum can be obtained up to an overall vertical shift.

Lastly, we would like to point out that the above algorithm and its justification also generalize if the product of determinants is replaced by an LLL-projection of the same product wave function. The EWFs should, in this case, be replaced with the corresponding LLL-projected counterparts.

\subsection{Approximation in LLL-projection} \label{LLLprojection}
The algorithm described in Sec.~\ref{sec:lzaBlocks} is exact up to statistical uncertainty from the Monte Carlo estimations of the matrix elements. The main set of results presented in this work is for the unprojected parton states. For these states, the algorithm provides a faithful estimator for the entanglement spectra. 
However, approximations are needed when calculating the spectra for the projected parton states. In this section, we describe the approximate projection which we used to obtain the LLL EWFs.

In this work we consider the ES of the projected wave function of the following form
\begin{gather}
    \psi^{2^{2}1^{4}}_{1/5}{=}\mathcal{P}_{\rm LLL}\ [\phi_1^2\phi_2]^2 {\sim}\left ( \mathcal{P}_{\rm LLL}\ [\phi_1^2\phi_2] \right)^{2}{\equiv}[\psi^{\rm Jain}_{2/5}]^{2}.
    \label{eq: parton_221111}
\end{gather}
The wave functions on either side of the $\sim$ sign in Eq.~\eqref{eq: parton_221111} differ in the details of how the LLL-projection is implemented. We expect such details to not affect the universality class of the phase described by the wave functions~\cite{Balram15a, Balram16b}. Only the form given on the rightmost side of Eq.~\eqref{eq: parton_221111} is amenable to a numerical evaluation for large system sizes and thus this is the form that will be used in our work.

Projecting state to LLL in the disc geometry is equivalent to replacing any $\bar{z}$ by $2\partial_{z}$ in the Slater determinant after taking all the terms with $z$ to the right hand side of the $\bar{z}$ (the derivatives do not act on the Gaussian part of the wave function)~\cite{Girvin84b}, i.e., 
\begin{equation}
\psi^{2^{2}1^{4}}_{1/5} =  \sbkt{\phi_2\rbkt{\{z,\bar{z}\rightarrow 2\partial_{z}\}}\phi_1^2}^{2}.
\end{equation}
Same replacements need to be applied in the EWFs to obtain their LLL-projections. Getting the exact analytical form of the LLL-projected state is not feasible beyond the case of $\sim 10$ particles. To make progress, we consider an approximate projection scheme where we replace $\bar{z}_i$ by the following.
\begin{equation}
\bar{z}_i\rightarrow \sum_{j\neq i}^{N}\frac{2}{z_i-z_j}
\end{equation}
For the Jain composite wave functions, this is a highly reliable way to perform the LLL-projections~\cite{Jain07}. This has also been used in the case of parton wave functions previously~\cite{Balram18, Balram18a}. We employ the same approximation to the case of the EWFs here. This produces wave function in the LLL with correlations similar to that of the unprojected wave functions. Thus, in general, we have our approximate projected parton state as 
\begin{equation}
\psi^{n^{m}1^{m}}_{n/[m(n+1)]}=  \sbkt{\phi_1\times \phi_n\rbkt{\cbkt{z,\bar{z}_i\rightarrow \sum_{j\neq i}^{N}\frac{2}{z_i-z_j}}}}^m
\end{equation}
which is used for RSES calculation.

\subsection{Edge counting from edge current algebra representations}
The parton states considered in this work namely $[\phi_1^a \phi_n]^k$ have edge mode currents carrying a representation of the $\widehat{su}(n)_k\times {u}(1)$ algebra. Algorithms discussed in the previous sections calculate the entanglement spectra reflecting the edge spectra of the states. To compare the entanglement spectra with the counting expected from the current algebra, in this section, we summarize the construction of the representations of the $\widehat{su}(n)_k$ algebra, which has the form 
\begin{equation}
[J_l^a,J_m^b]=\imath \sum_c f^{abc}J_{l+m}^c + k l\delta_{l+m,0}\delta_{a,b}.
\end{equation}
Here the indices $l,m\in \mathbb{Z}$, $1\leq a,b\leq {\rm dim}({su}(n))$ and $f^{abc}$ are the structure factors for the ${su}(n)$ Lie algebra. 
We provide a simplified picture of the construction, presented in parallel with ideas from the more familiar case of the highest weight representations of the ${su}(2)$ algebra. Further details on these ideas can be found in Chapters 13 and 14 of Ref.~\onlinecite{DiFrancesco97}.

As in the case of ${su}(2)$, the highest weight representations are labeled by the quantum numbers of the highest weight state, and the remaining basis states in the representation can be obtained by repeated action of ladder operators. The dominant highest weight representations of $\widehat{su}(n)_k$ are labeled by the Dynkin labels $\vec{\mu}=[\mu^{(0)},\mu^{(1)},\dots,\mu^{(r)}]$ of the highest weight state $|{\rm hw}\rangle$. Here $r$ is the dimension of the maximal commuting sub-algebra of $\widehat{su}(n)_k$, and $\mu^{(i)}$'s are non-negative integers that add up to $k$. Thus the highest weight representations of $\widehat{su}(2)_2$ (for which $r=1$) are labeled by Dynkin labels $[1,1]$, $[2,0]$ and $[0,2]$.

Having discussed the highest weight states, we now discuss the ladder operators. The ladder operators arise from identifying linear combinations of $\{J_n^a\}$ that form ${su}(2)$ sub-algebras of $\widehat{su}(n)_k$.
There are $r+1$ independent ladder operators $\{E_0,E_1\dots E_r\}$ that can be constructed (See Sec 13.1.3 and 14.1.4 of Ref.~\onlinecite{DiFrancesco97}). An infinite tower of basis states is generated by the ladder operators acting on the highest weight state. The action of a ladder operator on a basis state reduces the Dynkin label of the state by a fixed integral vector; each ladder operator $E_i$ is labeled by this integral vector $\vec{\alpha}_i$. These integral directions are determined by details of the algebra encoded in $f^{abc}$. For $\widehat{su}(2)_k$ these are given by $\vec{\alpha}_0=[2,-2]$ and $\vec{\alpha}_1=[-2,2]$. (See Sec. 14.1.7 of Ref.~\onlinecite{DiFrancesco97}). 

Repeated action of $E_i$ on $|{\rm hw}\rangle\equiv |\vec\mu\rangle$ creates an ${su}(2)$ multiplet. The dimension of this multiplet is $\mu^{(i)}+1$; thus each $E_i$ can act on the state $\mu^{(i)}$ times producing basis states $|\vec{\mu}\rangle,E_i^1|\vec{\mu}\rangle,E_i^2|\vec{\mu}\rangle,\dots,E_i^{\mu^{i}}|\vec{\mu}\rangle  $ with Dynkin labels $\vec{\mu},\vec{\mu}-\vec{\alpha}_i,\vec{\mu}-2\vec{\alpha}_i\dots \vec{\mu}-\mu^{(i)}\vec{\alpha}_i$. Any state $|\vec{\lambda}\rangle$ obtained by action of one of the ladder operators, say $E_i$, on any basis state is a highest weight state for the remaining $r$ ladder operators $E_j$, ($j\neq i$), and the corresponding multiplets have dimensions $\mu^{(j)}+1$ .

\begin{figure}[h]
	\includegraphics[width=\columnwidth]{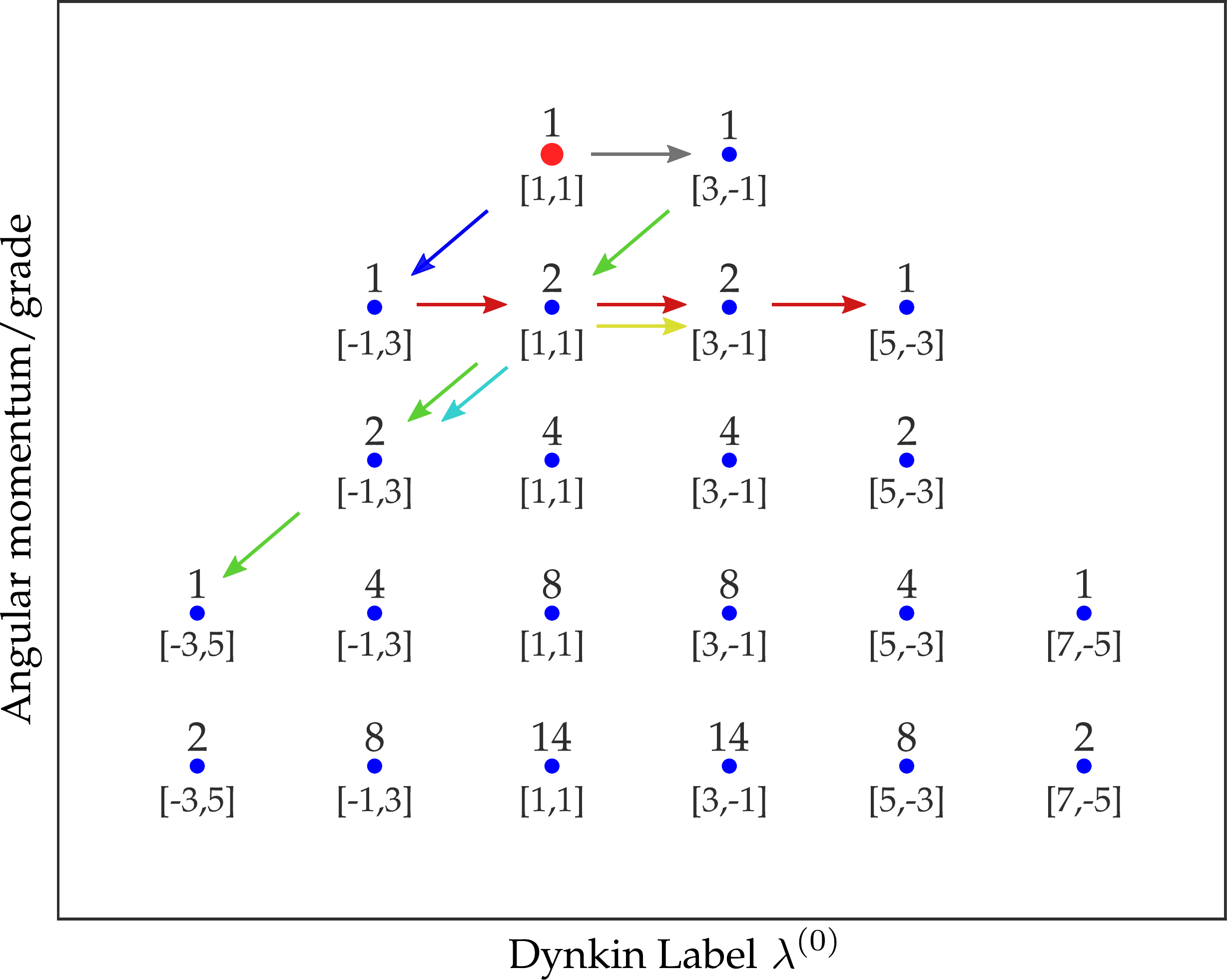}
	\caption{Representation of $\widehat{su}(2)_2$ generated from the highest weight state with Dynkin label $[1,1]$ (red dot). Other states in the representation are generated by action of ladder operators $E_0$ and $E_1$. Action of $E_0$ on any state is represented by a horizontal arrow and the diagonal arrows show the action of $E_1$. Arrows with same colors are used to illustrate successive application of either ladder operators to form an ${su}(2)$ multiplet.}
	\label{fig:example_weight_space}
\end{figure}

We illustrate this for the special case of $\widehat{su}(2)_2$. The ideas are summarized in Fig.~\ref{fig:example_weight_space} where we show the representation generated from the highest weight state $\vec{\mu}=[1,1]$ shown in the figure with a red dot. The action of $E_1$ is indicated by a horizontal arrow. This generates an ${su}(2)$ multiplet of dimension $\mu^{(1)}+1=2$. The action of the $E_0$ ladder operator is indicated by a diagonal arrow going down to the lower left. This creates a multiplet of dimension $\mu^{(0)}+1=2$.

\begin{figure*}
	\includegraphics[width=\textwidth]{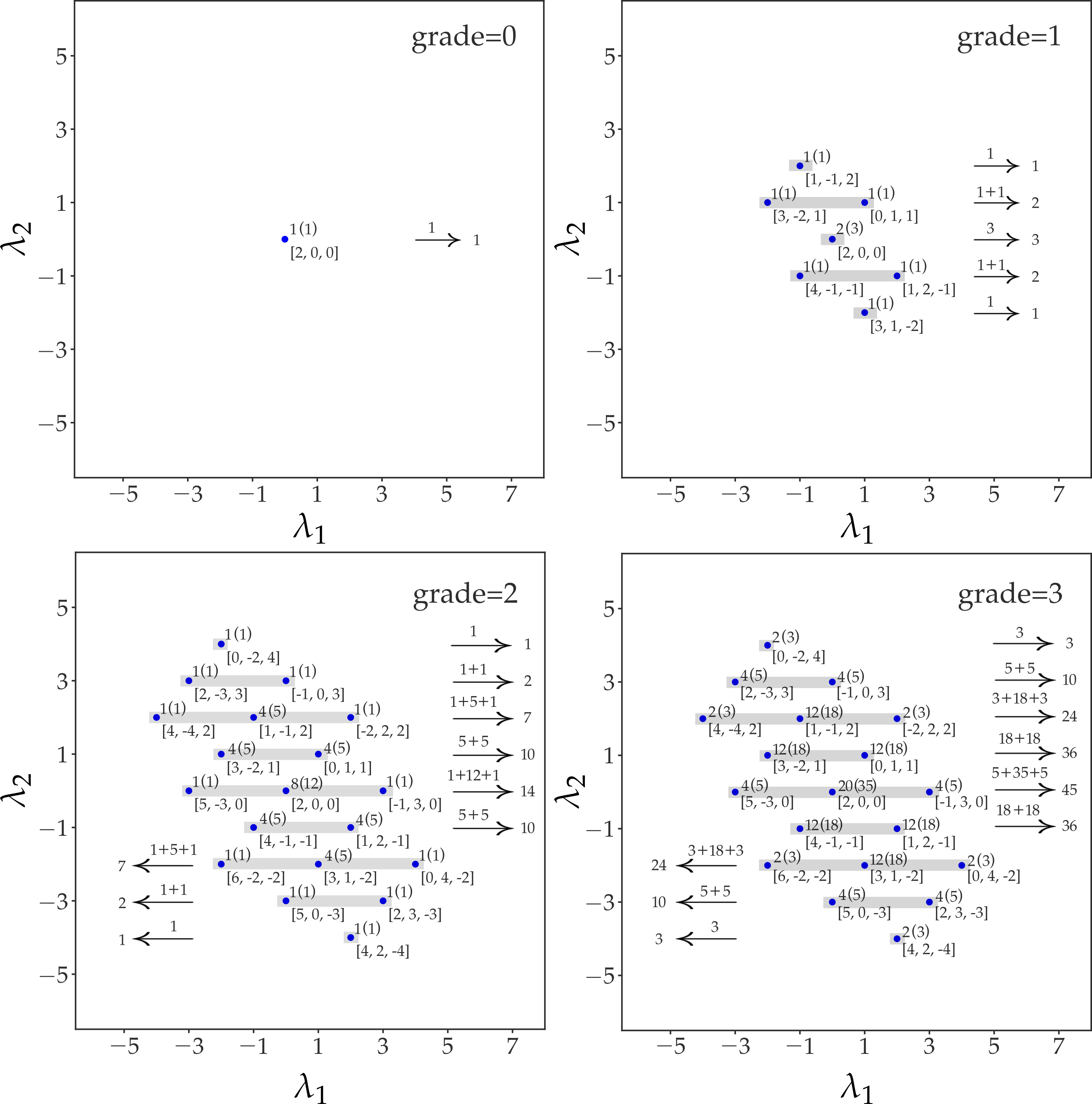}
	\caption{Representation of $\widehat{su}(3)_2$ characterized by the highest weight state with Dynkin label $[2,0,0]$. In each panel, which is labeled by the grade quantum number, action of $E_1$ and $E_2$ changes the Dynkin label without changing the grade. Highest weight state with Dynkin label $[2,0,0]$ has $\lambda_1,\lambda_2=0$, hence action of $E_1$ and $E_2$ annihilates the state. Therefore only one state exists for $\text{grade}=0$ which is represented by the blue marker in the top left panel. The number above each marker represents the multiplicity of the state whereas the number inside the parenthesis corresponds to the same after adding $1$ boson. Since $\lambda_0=2$, $E_0$ can act twice to produce states with Dynkin label $[0,1,1]$ and $[-2,2,2]$ which have grade $1$ and $2$ respectively. In the top right panel, we get six new Dynkin label after $E_1$ and $E_2$ act on $[0,1,1]$. 
		If we think of grade as $L_z^{A}$ quantum number in RSES, representation in each panel can be mapped to RSES in different $L_z^{A}$ sectors. The arrows give combined multiplicities (with $1$ boson) of states spanned by the grey boxes.  It turns out that these combined multiplicities are one-to-one mapped with the counting in $\phi_3^2$ RSES, shown in Fig.~\ref{fig:phi32ES}  (middle panel).
	}
	\label{fig:su32_200}
\end{figure*}

The state $E_1|\vec{\mu}\rangle$ is a highest weight state for $E_0$. It has a Dynkin label $[3,-1]$ and thus is the highest weight state of a dimension $4$ ${su}(2)$ multiplet generated by $E_0$ (indicated by the sequence of states connected by the green arrows). The state $E_0|\vec{\mu}\rangle$ has a Dynkin label $[-1,3]$ and is the highest weight state of a dimension $4$ ${su}(2)$ multiplet generated by $E_1$ (indicated by the sequence of states connected by the red arrows). 
The Dynkin label $[1,1]$ can be created by two paths in the diagram representing the states $E_1E_0|\vec{\mu}\rangle$ and $E_0E_1|\vec{\mu}\rangle $. Whether they are distinct states or linearly dependent is decided by the commutation relation between $E_1$ and $E_0$, encoded in the structure constants $f^{abc}$. In general, when a Dynkin label can be arrived at by multiple paths in the diagram, the multiplicity of that Dynkin label is determined by the number of linearly dependent combinations of $E_i$'s that reach there starting from $|\vec{\mu}\rangle$, which in turn is determined by the commutation relations between the ladder operators. 

Getting the multiplicities, taking into account all commutation relations can be difficult; thankfully the multiplicities can be obtained directly using the Freudenthal Recursion formula (See Eq 14.136 of Ref.~\onlinecite{DiFrancesco97}). The formula allowed us to evaluate the multiplicities using simple computational combinatorics. The multiplicities are shown in Fig.~\ref{fig:example_weight_space} as an integer above the dot. 

The coordinates of the dots in the Fig.~\ref{fig:example_weight_space} are chosen as follows. The $x$-axis represents $\lambda^{(0)}$ (note that the Dynkin label $\lambda^{(1)}$ is automatically determined as $k-\lambda^{(0)}$) and the $y$ coordinate called the ``grade" is determined by the number of actions of $E_0$ required to reach the state. We empirically identify the $E_0$ with angular momentum raising operator in the edge spectrum. Thus the $y$-coordinate is identified as angular momentum in the spectrum, which means the total number of states at angular momenta $0$, $1$, $2$ are $2$, $6$, $12$ respectively.

The edge current algebra is $\widehat{su}(n)_k\times {u}(1)$. The multiplicities at the angular momenta $L_{z}=0,1,2\dots$ for the $u_1$ boson is given by the integer partitions of $L_z$, namely $1,1,2,3,5\dots$ (See Sec 14.4.3 of Ref.~\onlinecite{DiFrancesco97}). The total multiplicity is given by the convolution $\mathcal{M}(L)=\sum_{m=0}^M \mathcal{M}_{\widehat{su}(n)_k}(m) \mathcal{M}_{{u}(1)}(L-m) $ considering all possible ways of partitioning the angular momenta. 
The representation of $\widehat{su}(2)_2$ and $\widehat{su}(2)_3$ are given in  Fig.~\ref{afig::su22All} and \ref{afig::su23All} respectively. In the figure, the multiplicity $\mathcal{M}_{\widehat{su}(n)_k}$ for each Dynkin label is shown above the dot whereas the full multiplicity $\mathcal{M}(L)$ is given in the parenthesis. Fig.~\ref{fig:su32_200} shows the representation of $\widehat{su}(3)_2$ for highest weight state given by Dynkin label $[2,0,0]$. Other highest weight representations are presented in the appendix (see Figs.~\ref{fig:su32_002} and \ref{fig:su32_011} of App.~\ref{app: rep_su2_level2}).


\section{Results}
\label{sec: results}
In this section, we present the numerically obtained RSES for three different parton states as well as for their quasihole excitations. We compare the structure of RSES with the representations of the corresponding edge current algebra.It is found that the RSES contains multiple distinct branches. This is true for Jain CF states as well where it has been found that the different branches can be associated with the different number of composite fermions occupying the different Lambda levels~\cite{Sreejith11c}.We can attribute a similar origin to the different branches in RSES of these partons as well. For this, we find the number of linearly independent edge states after fixing the number of particles in the different LLs in each parton.

\subsection{Real-space entanglement spectra of parton states}
\label{sec:RSESParton}
\begin{figure}[h!]
	\includegraphics[width=\columnwidth]{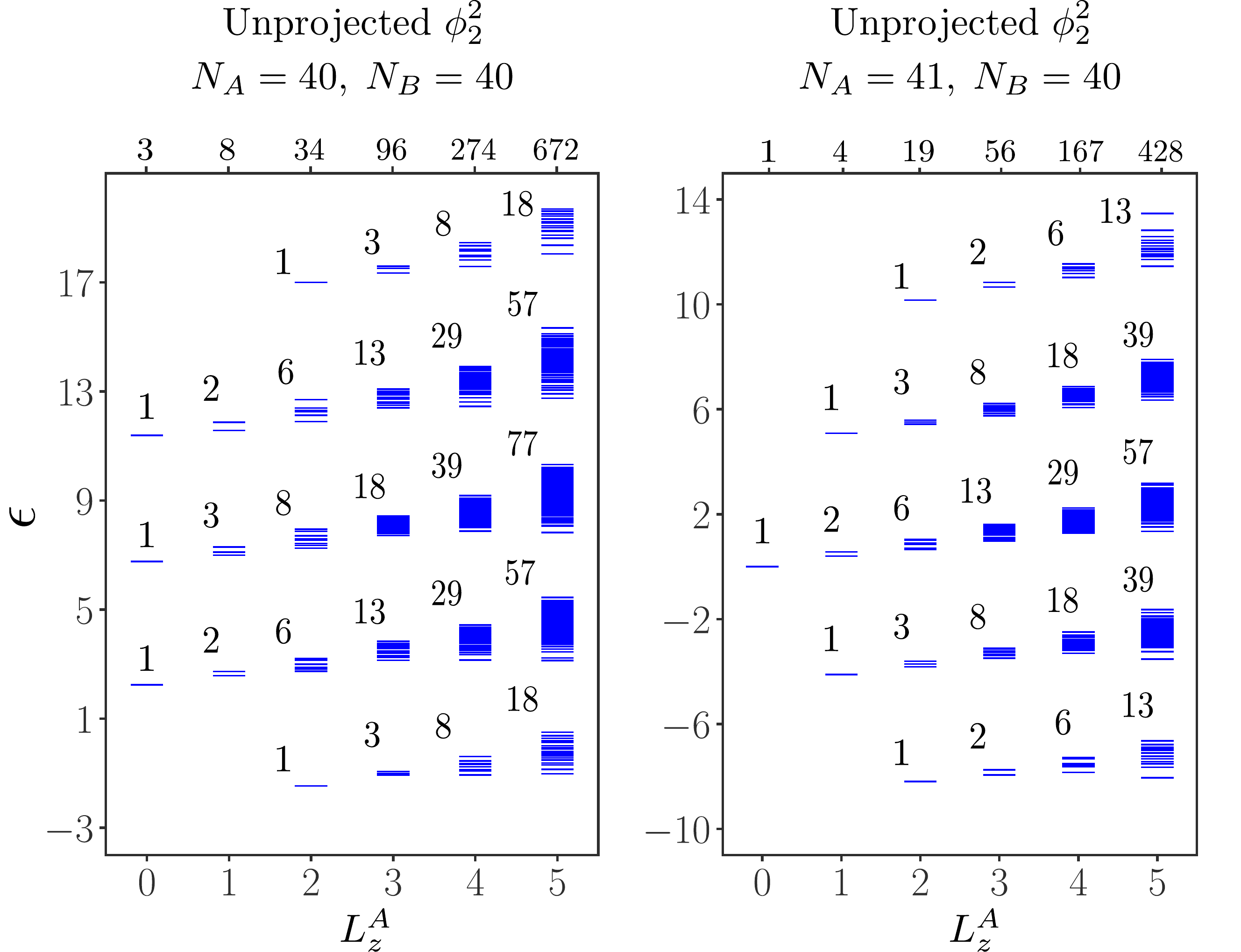}
	\caption{RSES of the parton wave function unprojected $\phi^{2}_{2}$. The two different  panels show the RSES in two different blocks of $\rho_A$ corresponding to $N_{A}=N_{B}=40$ (left) and $N_{A}=41,$ $N_{B}$ $=40$ (right). Entanglement energies $\epsilon$  are the negative logarithm of eigenvalues corresponding to $L_z^{A}$-blocks of $\rho_A$ in a given $(N_A,N_B)$-block. Each marker thus is labeled by its $L_z^{A}$ quantum  number on $x$-axis which is shifted such that  $L_{z,{\rm min}}^{A}=0$. The number next to clusters of markers represents the number of eigenvalues in that cluster. Total number of distinct EWFs in any given $L_z^{A}$-block is shown just above the top axis.}
	\label{fig:phi22ES}
\end{figure}

Fig.~\ref{fig:phi22ES} shows the RSES of the $\phi_2^2$ state across a rotationally symmetric cut (schematically shown in Fig.~\ref{fig:schematic_realspace_cut}). The left panel shows the RSES corresponding to a block of the reduced density matrix with $N_A=40$ and the right panel is for the block with $N_A=41$. RSES is qualitatively unchanged when $N_B$ is varied without changing $N_A$.
For other nearby $N_A$ sectors, the RSES is qualitatively the same as the ones shown here - depending on the parity (odd/even) of $N_A$.

The density matrix $\rho_{A}$ within an $N_A$ block is itself block diagonal in angular momentum sectors. 
The $x$-axis shows the angular momentum eigenvalues ($L_z^A$) of different blocks. The $y$-axis shows the (negative logarithm of) eigenvalues of that block of the reduced density matrix.

\begin{figure}[h!]
	\includegraphics[width=\columnwidth]{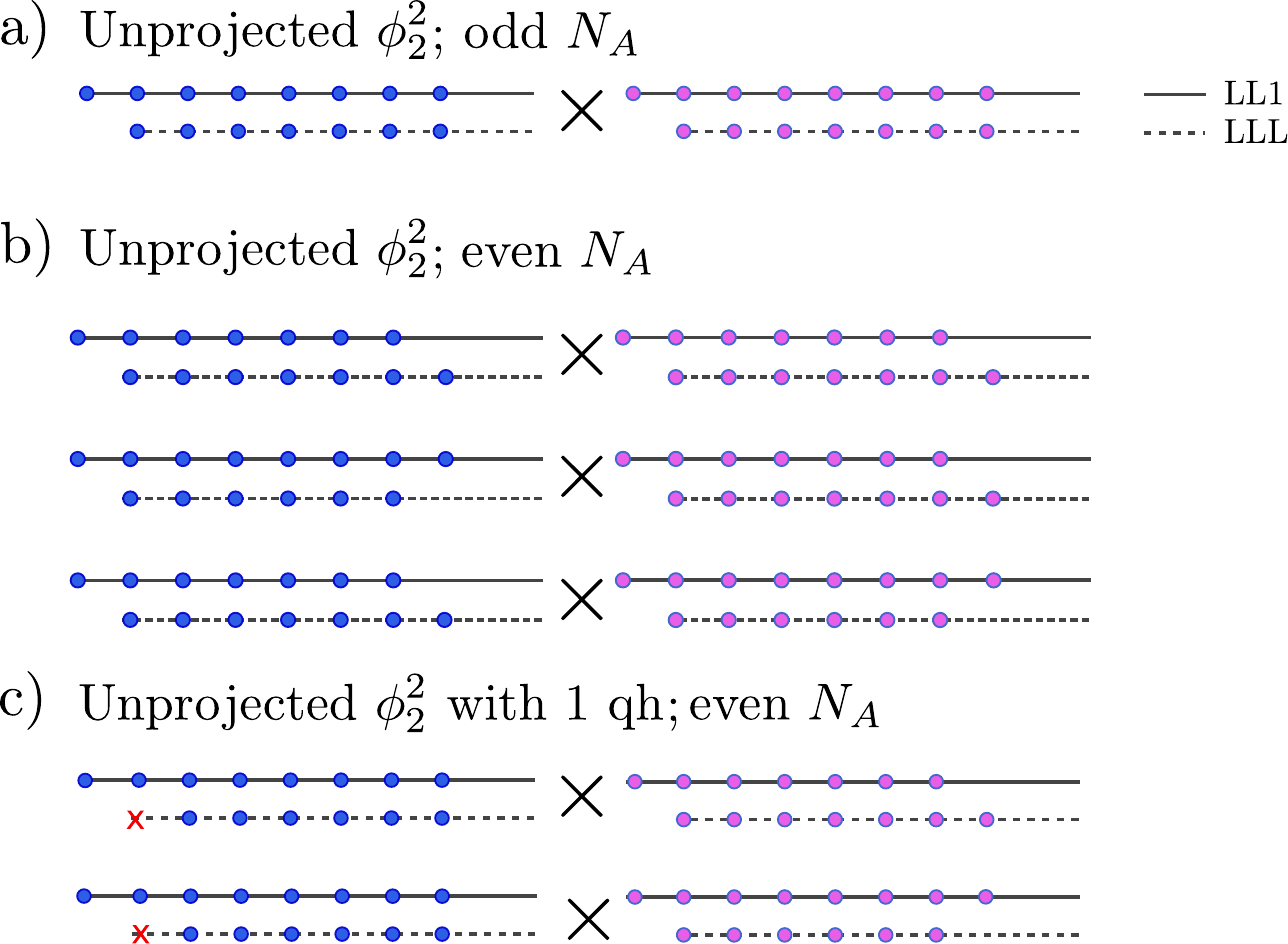}
	\caption{Schematic diagram showing all compact LL configurations for unprojected $\phi_2^2$ in even/odd $N_A$ sectors. Different partons are represented with different colors. For (a) odd $N_A$, there is only one configuration with $L^A_{z,{\rm min}}$ whereas for (b) even $N_A$, three distinct LL configurations are possible. We also study RSES for parton states with one quasihole excitation. Panel (c) shows two possible compact LL configurations for unprojected $\phi_2^2$ with one quasihole in LLL, when $N_A$ is even. }
	\label{fig:minimumEWF}
\end{figure}

In the expansion shown in Eq.~\eqref{eq: DetToEntanglementWFs}, the EWF $\xi_\lambda^A$ with the smallest angular momentum corresponds to that in which all the $N_A$ particles in each parton compactly occupy the smallest single-particle angular momentum states. All possible compact LL configurations for $\phi^2_2$ are shown in Fig.~\ref{fig:minimumEWF}.
In the case of $\phi^2_2$ with odd $N_A$, the lowest angular momentum is obtained when both the partons have $(N_A-1)/2$ particles in the LLL and $(N_A+1)/2$ in the 2nd LL. This is the only state with this angular momentum. When $N_A$ is even, there are three distinct EWFs, all with an angular momentum equal to the lowest angular momentum.
There are no EWFs with $L_z^A$ below these angular momenta, and hence the RSES contains no states for $L_z^A<L_{z,{\rm min}}^A$.
The $L_z^A$ values across the $x$-axis in these figures are relative to this minimum angular momentum $L_{z{\rm min}}^{A}$.

The entanglement energies in the RSES appear in several nearly-degenerate clusters. In Fig.~\ref{fig:phi22ES}, the number of entanglement energies in a cluster is shown right next to the cluster.
The total number of distinct EWFs with a given $L_z^A$ (and $N_A$) obtained by simple enumeration (as described in Sec. ~\ref{sec:lzaBlocks}) increases quite rapidly with $L_z^A$. This number is shown just above the top axis of each panel.
The total number of entanglement energies in each $L_z^A$ sector is much less than the number of EWFs. This is due to the extensive number of linear dependencies between the EWFs. As discussed in Sec.~\ref{sec:lzaBlocks}, these linear dependencies are reflected as $0$ eigenvalues of the matrix $M$.\label{key}

In general, the total number of the EWFs, total number of linear dependencies, and the total number of entanglement energies in each $L_z^A$ sector grow rapidly with $L_z^A$. Due to the proliferation of EWFs, reliable numerical calculation of the matrix $M$ is possible only up to $L_z^A \sim 5$. These sectors were sufficient to make a comparison with the representations of the current algebra.
Quad-precision arithmetic in C codes reliably estimates $M$ in systems as large as $80$ particles. Statistical uncertainties in the estimated eigenvalues can be inferred by considering the results of many independent Monte Carlo trials. These error bars are typically comparable to the width of the marker and therefore are not shown.

The $y$-axes of the RSES plots are shifted by an arbitrary constant.
This is because the matrix $M$ and therefore its eigenvalues can be calculated only up to a proportionality constant in the Monte Carlo techniques (see Sec \ref{sec:lzaBlocks}). This factor depends only on the Monte Carlo sampling distribution. Since the same distribution is used for all $L_z^A$ sectors within a single plot, the different $L_z^A$ sectors are shifted by the same amount. 
For the calculations presented in each plot, the Monte Carlo sampling wave function was chosen to be equal to the square of the absolute value of one of the EWFs with angular momentum $L_{z,{\rm min}}^A$. Since the sampling function is different across different plots (for instance the two panels in Fig.~\ref{fig:phi22ES}), the ES in different plots are shifted by different constants.

The radius of the circular cut, (Fig. ~\ref{fig:schematic_realspace_cut}) was chosen to be approximately equal to the radius of EWF with the most compact configuration i.e. with a total angular momentum $L_{z,{\rm min}}^A$. For example, in the case of $\phi^2_2$, the largest single-particle momenta of the most compact configuration is approximately $N_A$. Hence the radius of the cut was chosen as $\sqrt{2 N_A}$ (in magnetic length units).

\textit{RSES of $\phi_2^2$:}  Fig.~\ref{fig:phi22ES} shows the RSES of unprojected $\phi_2^2$ where the left and right panels correspond to even and odd $N_A$-blocks respectively.  Across different $L_z^A$ sectors, the clusters in entanglement energies appear as distinct branches. Different arrangement of these branches give the RSES a particular structure. RSES of unprojected $\phi_2^2$ for even and odd $N_A$ consist of different arrangements of two distinct branches ($1,2,6,13,\dots$) and ($1,3,8,18\dots$) as shown in Fig.~\ref{fig:phi22ES}.  The counting and structure of RSES for even $N_A$ exactly matches to the $\widehat{su}(2)_2\times {u}(1)$ representation corresponding to Dynkin label $[0,2]$ whereas for odd $N_A$, it matches with representation corresponding to Dynkin label $[2,0]$ (left and middle panels in Fig.~\ref{afig::su22All}).

\begin{figure}[h!]
\includegraphics[width=\columnwidth]{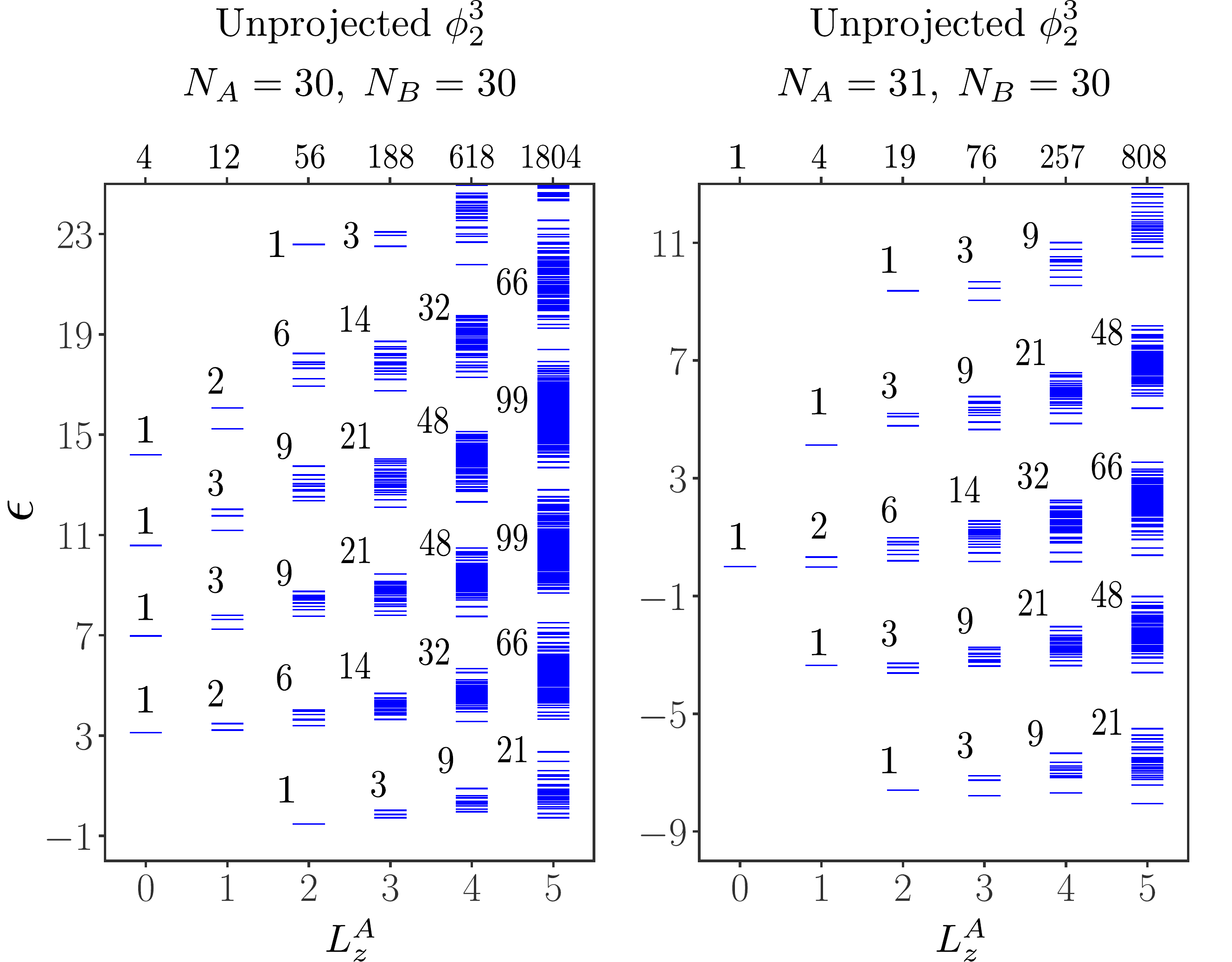}
\caption{RSES of unprojected $\phi^{3}_{2}$ for  $N_{A}=N_{B}=30$ (left) and $N_{A}=31,N_{B}=30$ (right) blocks of $\rho_A$. Number of distinct EWFs for $L_z^A$-sectors are much larger than that for $\phi_2^2$ parton state, and hence spread in clusters in more due to numerical errors.\label{fig:phi23ES}}
\end{figure}

\textit{RSES of $\phi_2^3$:}  The RSES of unprojected $\phi_2^3$ is shown in Fig.~\ref{fig:phi23ES} with the left panel showing the spectrum for $(N_A=30,N_B=30)$ sector and right panel for $(N_A=31,N_B=30)$. Similar to the case of $\phi_2^2$, we find that the RSES arrangement of unprojected $\phi_2^3$ only depends on the parity (even/odd) of $N_A$. These arrangement consists of two distinct branches which have counting $(1,2,6,14,32,66,\dots)$ and $(1,3,9,21,48,99,\dots)$. We find that the RSES arrangement for even $N_A$ exactly matches the highest weight representation of $\widehat{su}(2)_3\times {u}(1)$ corresponding to Dynkin label $[0,3]$ and the representation corresponding to Dynkin label $[3,0]$ matches spectrum for odd $N_A$ (Fig.~\ref{afig::su23All}).

\begin{figure}[h!]
	\includegraphics[width=\columnwidth]{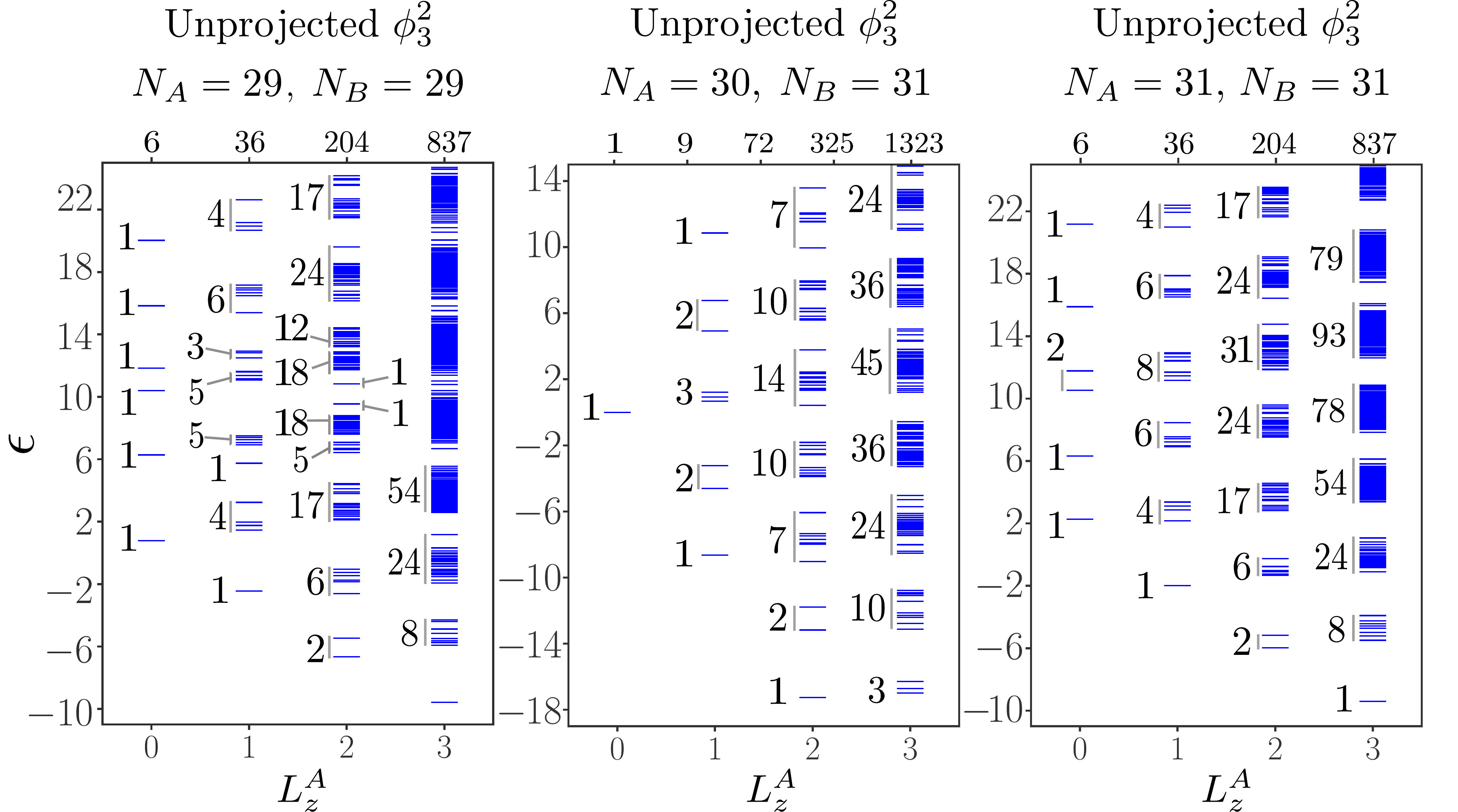}
	\caption{ RSES of unprojected $\phi^{2}_{3}$ for three blocks of $\rho_A$ corresponding to $N_{A}=N_{B}=29$ (left), $N_{A}=30,N_{B}=31$ (mid) and $N_{A}=31,N_{B}=31$ (right). As multiple branches emerge in close proximity to each other,  we use a vertical (grey) line to show which clusters are clubbed together. Total number of states in these collections are shown next to the vertical lines.
		\label{fig:phi32ES}}
\end{figure}

\paragraph*{RSES of $\phi_3^2$:} Fig.~\ref{fig:phi32ES} shows the RSES of unprojected $\phi_3^2$ where the left, middle and right panels show the RSES for $(N_A=29,N_B=29)$, $(N_A=30,N_B=31)$ and $(N_A=31,N_B=31)$ blocks respectively.
Three different branches are present in the RSES corresponding to  $(N_A=29,N_B=29)$ and $(N_A=31,N_B=31)$-sectors, with counting $(2,8,31,93,..)$, $(1,4,17,54,..)$ and $(1,6,24,78,..)$. For the case of $(N_A=30,N_B=31)$, we find branches with counting $(1,3,14,45,..)$, $(2,10,36,..)$, and $(1,7,24,..)$. We use a grey vertical line in Fig.~\ref{fig:phi32ES}  to represent which eigenvalues are clubbed together in the counting.

Unlike the case of $\phi_2^2$ and $\phi_2^3$, comparing RSES of $\phi_3^2$ to $\widehat{su}(3)_2\times {u}(1)$ representation is not straightforward. Let's take an example of RSES for the $(N_A=30,N_B=31)$ sector (middle panel of Fig.~\ref{fig:phi32ES}). Total number of states at each angular momentum in the RSES match the total number of states at the corresponding grade in Fig.~\ref{fig:su32_200}. We empirically find that the counting in the branches in the RSES is the sum total of counting along a horizontal line in Fig.~\ref{fig:su32_200} (shaded grey boxes). In this sense, the spectrum for $N_A=30$ matches the $[2,0,0]$ representation of the $\widehat{su}(3)_2\times {u}(1)$ algebra. Other highest weight representations of $\widehat{su}(3)_2\times {u}(1)$ are given in App.~\ref{app: rep_su2_level2} (see Figs.~\ref{fig:su32_002}-\ref{fig:su32_011}). We find that the $N_A=29$ and $N_A=31$ RSES matches highest weight representations for $[0,0,2]$ and $[0,2,0]$ Dynkin labels respectively, shown in Fig.~\ref{fig:su32_002}. In all the cases that we tested, we find that the aggregate number of states along a line matches exactly with the spectrum obtained from the current algebra.

The finer structures in each cluster too resemble the counting in the states of individual Dynkin labels. We point this out in the RSES of $N_A=29$ shown in the left panel of Fig.~\ref{fig:phi32ES}. For instance, at $L_{z}=1$, we find seven clusters with counting $(1,4,6,8,6,4,1)$, which can be associated with the total number of states along horizontal lines in the top right panel (grade 1) of Fig.~\ref{fig:su32_002}. The number $4$ is obtained by adding three states with Dynkin labels $[0,0,2]$ and one state with Dynkin label $[3,-3,2]$. This 3+1 structure is reflected in the RSES as indicated in the left panel of Fig.~\ref{fig:phi32ES}. 

\begin{figure}[h!]
        \includegraphics[width=\linewidth]{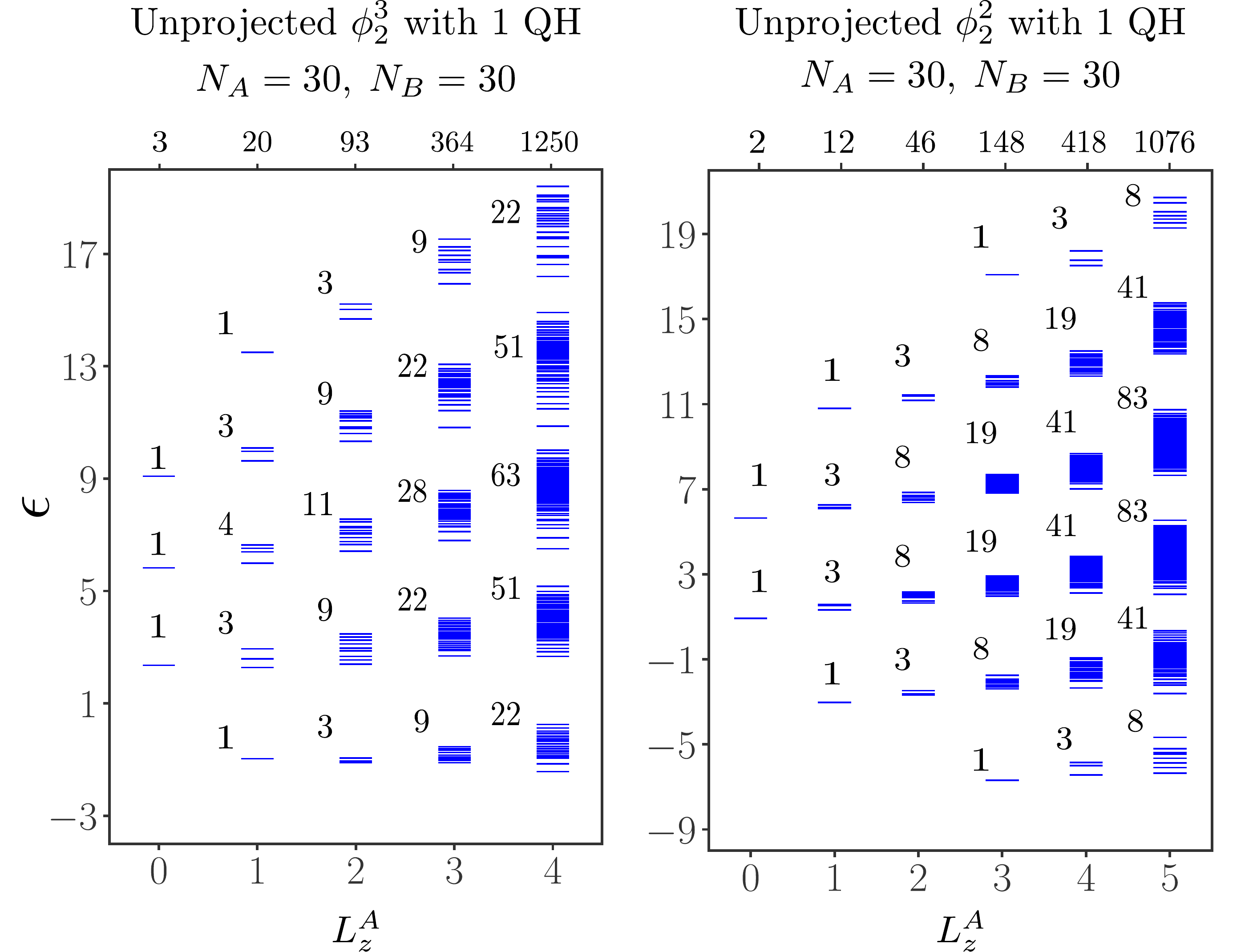}
    \caption{RSES of $\phi^{3}_{2}$ (left)  and  $\phi^{2}_{2}$ (right) where both parton states have a single quasihole placed at $m=0$ angular momentum state of LLL. Both panels show RSES for $N_{A}=N_{B}=30$ block of $\rho_A$. }
    \label{fig:1QH_ES}
\end{figure}

\paragraph*{RSES of $\phi_2^2$ and $\phi_2^3$ with a quasihole:}  We also explore the RSES for $\phi_2^2$ and $\phi_2^3$ when $1$ quasihole is added to the system at the origin (Fig.~\ref{fig:1QH_ES}). Fig.~\ref{fig:minimumEWF} provides a schematic for compact LL occupation of $\phi_2^2$ when a quasihole is added in the LLL and $N_A$ is even. We found that RSES remains the same whether we put the quasihole in  LLL or the second LL (LL1).

Left panel of Fig.~\ref{fig:1QH_ES} shows the RSES for $\phi_2^2$ with a quasihole and $N_A=31$. We found that the same RSES is produced for both odd and even $N_A$.
RSES for $\phi_2^2$ with a quasihole exactly matches with the $[1,1]$ highest weight representation of $\widehat{su}(2)_2\times {u}(1)$ (middle panel in  Fig.~\ref{fig:phi22ES}).
RSES for $\phi_2^3$ with a quasihole, for $N_A=30$ (right panel in Fig.~\ref{fig:1QH_ES}) matches with with $[1,2]$ highest weight representation of the $\widehat{su}(2)_3\times {u}(1)$. RSES spectra changes for $N_A=31$ which maps exactly to $[2,1]$ highest weight representation of $\widehat{su}(2)_3\times {u}(1)$. The corresponding representations are shown in the bottom panels of Fig.~\ref{afig::su23All}

\subsection{RSES for the projected state}
Until now we discussed the RSES of unprojected parton states. A natural question is whether the agreement between the RSES and the representations of the edge currents continues to hold even after projection into the LLL. The algorithm that we have used for the calculation extends also to the case of the projected parton states. The calculations run slower due to the computational cost of projection. Moreover, an exact projection into the LLL is not possible except in very small systems. We instead rely on approximate projection as described in Sec. \ref{LLLprojection}. To enable the use of the projection scheme, we consider a projected state $\psi^{2^{2}1^{4}}_{1/5}$ [see Eq.~\eqref{eq: parton_221111}] which is expected to also show the same edge counting as $\phi_2^2$.

\begin{figure}[h!]
	\includegraphics[width=\columnwidth]{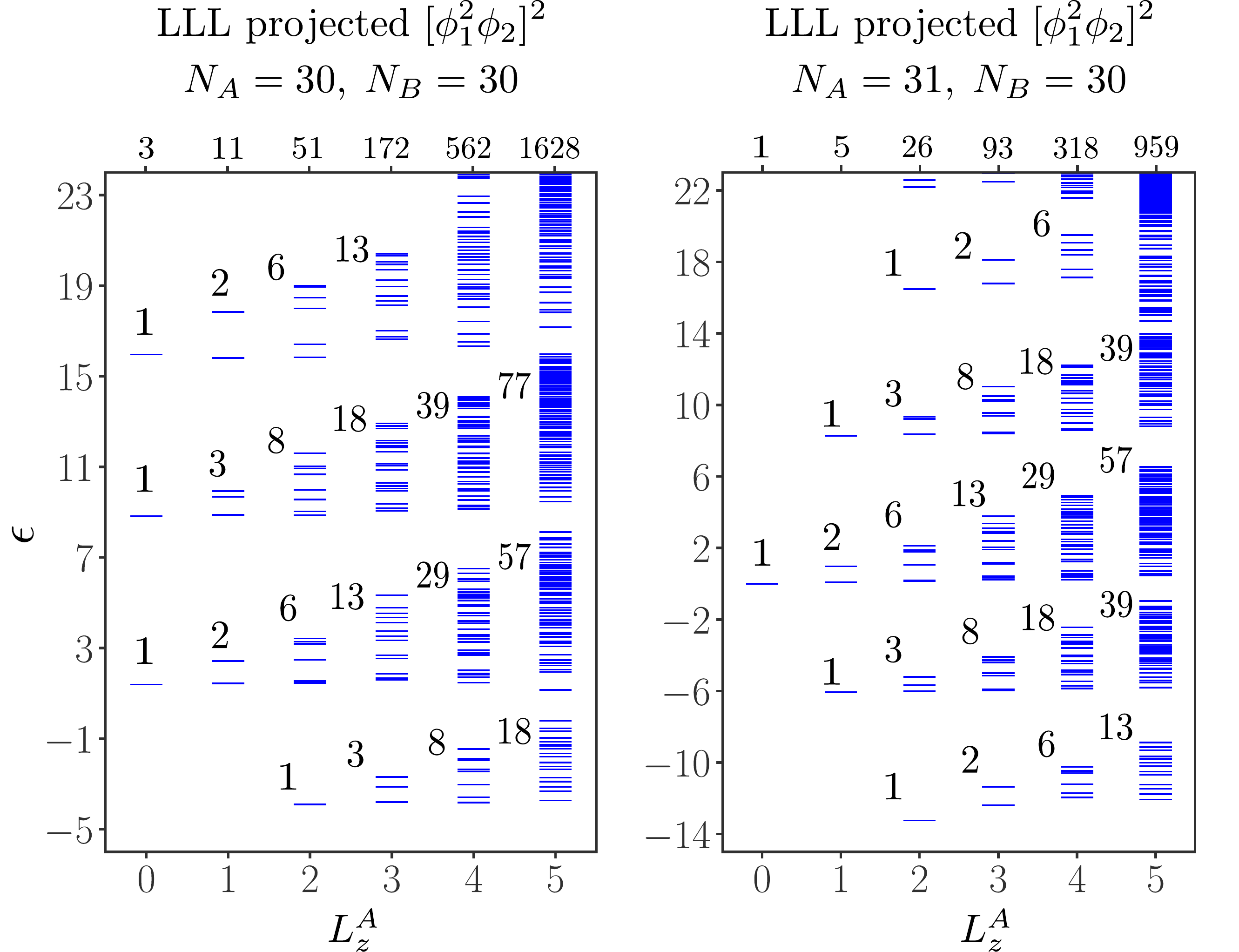}
	\caption{RSES of LLL-projected $[\phi_1^2\phi_{2}]^2$ for two blocks of $\rho_A$ corresponding to $N_{A}=N_{B}=30$ (left) and  $N_{A}=31,N_{B}=30$ (right). The projection is approximated as described in Sec \ref{LLLprojection}. Apart }
	\label{fig:proj_ES}
\end{figure}

We indeed find that even after the approximation, the RSES of LLL-projected $\psi^{2^{2}1^{4}}_{1/5}$ state has the same structure $\phi_2^2$ as shown in Fig.~\ref{fig:proj_ES}. Note that the number of the EWFs (shown above the panels) in each $L_z^A$ block is significantly larger than that for unprojected $\phi_2^2$ (Fig.~\ref{fig:phi22ES}). This is a result of the excitations associated with the additional Slater determinants $\phi_1$ in the wave function. We find a larger spread in the clusters. An exact match with RSES of unprojected $\phi_2^2$ suggests that multiplication with factors of $\phi_1$ and action of LLL-projection do not alter the topological features (fusion rules of the non-trivial anyons are identical) of a parton state. 

\subsection{Origin of the branches and edge overlap matrix}
The branches in the entanglement spectra appear in the case of the edge spectrum of composite fermion states as well \cite{Sreejith11c, Rodriguez13}. Each branch in that case could be associated with states with a particular occupancy of the composite fermion Landau levels. In this section, we ask whether a similar picture is true in the context of the parton states. 

The entanglement spectrum contains states with different occupancy of the Landau levels of each parton. A strategy to address the question could be to expand the Schmidt eigenstates in the EWFs. Then identify the number of particles in each LL in the basis states that contribute predominantly to each Schmidt eigenvector. This strategy does not yield a clear answer as the EWFs are not all linearly independent. 

We instead address the problem in the following way. We consider the states with a fixed number of particles in each LL of each parton. We consider the excitations at the edge of the system at each angular momentum. These are the same as the EWFs that we had constructed earlier but with a specific Landau level occupancy of each parton. We consider the overlap matrix of the edge states, $\mathcal{O}$ and plot the (negative logarithm of) eigenvalues for different $L_z$-blocks of $\mathcal{O}$. The number of dominant eigenvalues is a measure of the linearly independent states in the space of these states i.e the dimension of the space.

\begin{figure}[h!]\includegraphics[width=\columnwidth]{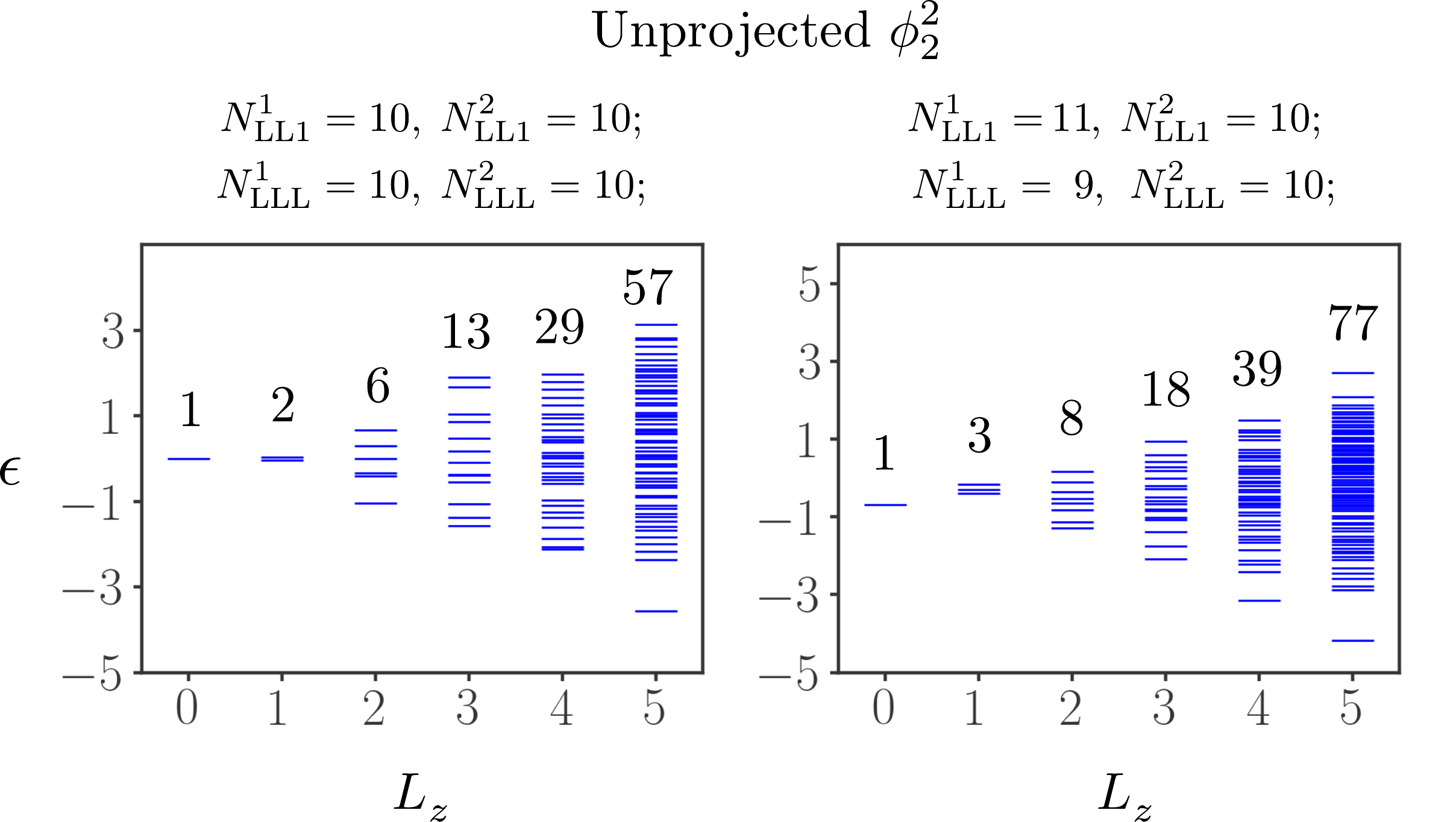}
	
	\caption{Dominant eigenvalues of the overlap matrix $\mathcal{O}$ for two different Landau level occupancies. Left panel shows the spectra  when both partons have identical LL occupation $(N_{\rm LLL},N_{\rm LL1})=(10,10)$. The right one is for when LL occupation of one of the partons is $(N_{\rm LLL}+1,N_{\rm LL1}-1)$ and the other one has a configuration $(N_{\rm LLL},N_{\rm LL1})$. $y$-axis is the negative logarithm of eigenvalues of different $L_z$-blocks of $\mathcal{O}$.}
	\label{fig:edge_phi22}
\end{figure}

\begin{figure}[h!]
	\includegraphics[width=\columnwidth]{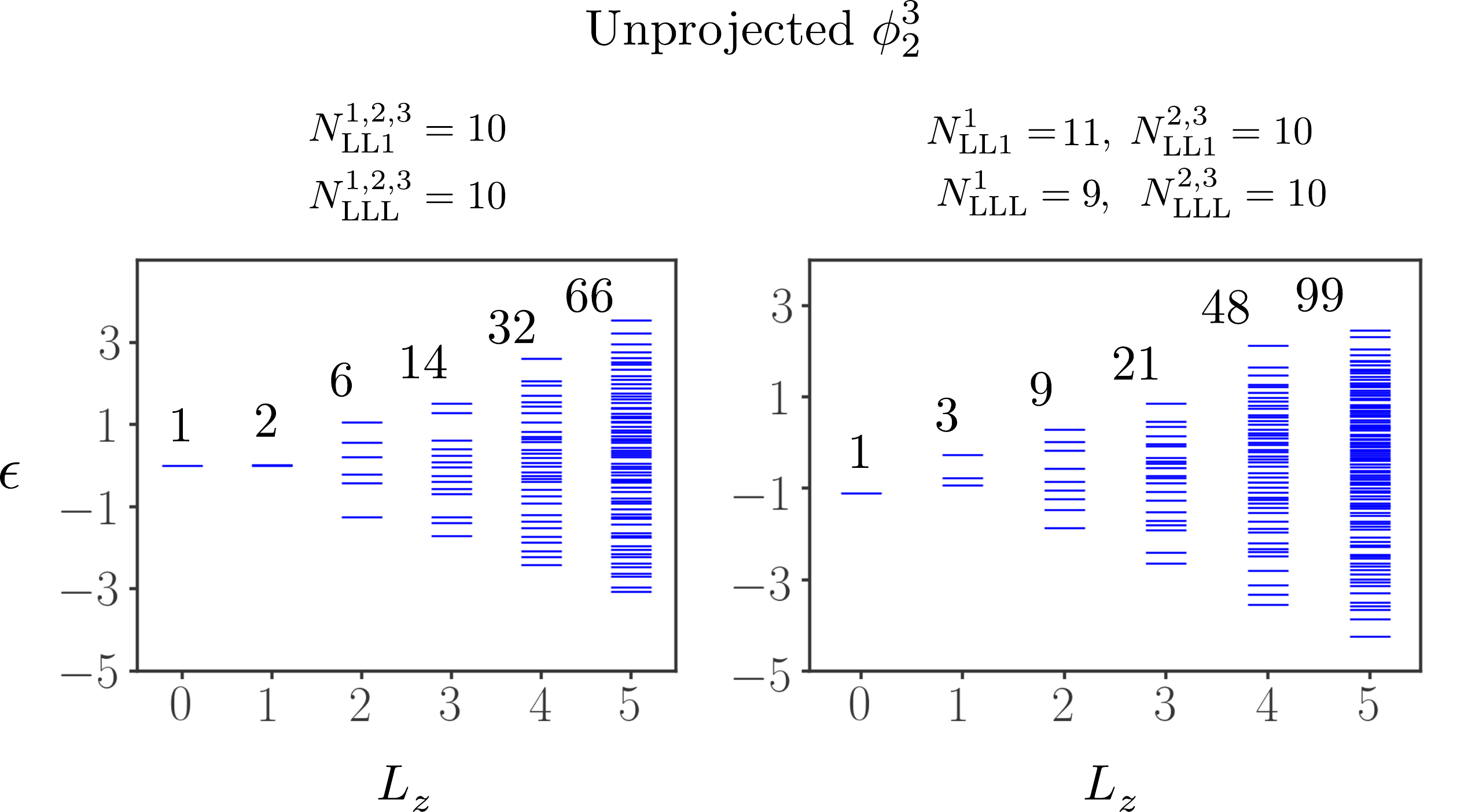}
	\caption{Dominant eigenvalues  of overlap matrix $\mathcal{O}$ of $\phi_2^3$ for different Landau level occupancies. Left panel shows spectra for a configuration when all three partons have identical LL occupation $(N_{\rm LLL},N_{\rm LL1})$.  The right one is for a configuration when two partons have same LL occupation $(N_{\rm LLL},N_{\rm LL1})$ and LL configutaion for the third parton  is given by $(N_{\rm LLL}+1,N_{\rm LL1}-1)$.  Two distinct counting we get in these  spectra match exactly to those of two distinct branches present in the RSES of $\phi_2^3$.}
	\label{fig::edge_phi23}		
\end{figure}

Interestingly, the number of dominant eigenvalues of  $\mathcal{O}$ at different $L_z$ values reproduce the same counting that we saw in the individual branches present in the RSES. The dominant eigenvalues as a function of momentum are plotted in Figs.~\ref{fig:edge_phi22} and \ref{fig::edge_phi23} for the case of $\phi_2^2$ and $\phi_2^3$ respectively, and further results are summarized in Tables \ref{tab1} and \ref{tab2} respectively.

\paragraph*{For $\phi_2^2$:} Fig.~\ref{fig:edge_phi22} shows dominant eigenvalues of the overlap matrix against $L_z$ for two different $(N_{\rm LLL},N_{\rm LL1})$-sectors. The left panel shows the case of $(N^{1,2}_{\rm LLL},N^{1,2}_{\rm LL1})= (10,10)$ (where superscript labels different partons). As shown in Table \ref{tab1}, as long as both the partons have same LL occupation, the counting remains the same. 
The right panel shows the case where $(N^{1}_{\rm LLL},N^{1}_{\rm LL1})= (11,9)$ and $(N^{2}_{\rm LLL},N^{2}_{\rm LL1})= (10,10)$. This produces a different branch. Similarly, we find that more generally whenever $N^{1}_{\rm LLL}=N^{2}_{\rm LLL}+1$ and  $N^{1}_{\rm LL1}=N^{2}_{\rm LL1}-1$, we get the same counting (see Table \ref{tab1}). The counting that we see in these two cases is in agreement with the individual branches seen in the ground state RSES shown in Fig.~\ref{fig:phi22ES}.

\begin{table}[h]
\begin{tabular}{ | P{3.5 cm}  | P{2.5cm} | } 
\hline
\small
    $(10,10)\times (10,10)$  &\multirow{4}{*}{1,2,6,13,29,57}  \\ \cline{1-1}
    $(11,9)\times (11,9)$&   \\ \cline{1-1}
    $(12,8)\times (12,8)$&   \\ \cline{1-1}
    $(11,10)\times (11,10)$&   \\ \hline

    $(11,9) \times (10,10)$ &\multirow{3}{*}{ 1,3,8,18,39,77}  \\ \cline{1-1}
    $(12,8) \times (11,9)$&  \\ \cline{1-1}
    $(12,9) \times (11,10)$&  \\ \hline
    
     $(10 , 10+ \textcolor{black}{q}) \times (10,10)$ & \multirow{3}{*}{ 1,3,8,19,41,83}  \\ \cline{1-1}
    $(11,9+ \textcolor{black}{q} ) \times (11,9) $ & \\ \cline{1-1}
    $(11,10+ \textcolor{black}{q}) \times (11,10)$ & \\ \hline
\end{tabular}
\caption{Different LL configurations for parton state $\phi_2^2$ given by  $(N^{1}_{\rm LLL},N^{1}_{\rm LL1})\times (N^{2}_{\rm LLL},N^{2}_{\rm LL1})$ along with counting corresponding to the dominant spectra of its overlap matrix $\mathcal{O}$. Here $N_{LLn}+q$ represents the quasihole addition to $n$th LL.}
\label{tab1}
\end{table}
If we perform a similar calculation of the eigenvalues of the overlap matrix for cases where there is a quasihole at the center, we obtain a counting that matches the individual branches of the RSES shown in the right panel in Fig.~\ref{fig:1QH_ES}). This counting is independent of the number of the particles in each parton Landau level (Table \ref{tab1}). This is consistent with the fact that only one type of branch appear in this RSES.

\paragraph*{For $\phi_2^3$:} Fig.~\ref{fig::edge_phi23} shows the dominant eigenvalues of the overlap matrix for $\phi^{3}_{2}$ in two different LL occupation sectors. Counting matches the different branches present in the $\phi_2^3$ RSES (Fig.~\ref{fig:phi23ES}). The spectra in the left panel corresponds to LL configuration where all partons have same LL occupation  $(N^{1,2,3}_{\rm LLL},N^{1,2,3}_{\rm LL1})=(10,10)$. We checked that the counting remains same as long as all three partons have same LL occupation (Table \ref{tab2}). Similarly, we get the second branch (right panel in Fig.~\ref{fig::edge_phi23}) in the RSES of $\phi^{3}_{2}$ when LL occupation is such that $N^1_{\rm LLL}=N^{2,3}_{\rm LLL}+1$ and $N^1_{\rm LL1}=N^{2,3}_{\rm LL1}-1$. 

RSES of $\phi^{3}_{2}$ with the quasihole has two distinct branches given by counting $(1,3,9,22,51,\dots)$ and $(1,4,11,28,63,\dots)$ as shown in Fig.~\ref{fig:1QH_ES}. We get the first branch from the dominant spectra corresponding to the LL configuration in which all three partons have the same LL occupation but one parton has a quasihole in LL1 (see Table \ref{tab2}). We get the second branch in configurations where $N^1_{\rm LLL}=N^{2,3}_{\rm LLL}+1$ and $N^1_{\rm LL1}=N^{2,3}_{\rm LL1}-1$ and the quasihole is in one of the latter (indexed by $2$,\,$3$) partons.

\begin{table}[h]
	\begin{tabular}{ | P{4.5 cm} |  P{2.5cm} | } 
		\hline
		$(10,10)^{3}$ & \multirow{3}{*}{ 1,2,6,14,32,66}  \\ \cline{1-1}
		$(11,9)^{3}$&  \\ \cline{1-1}
		$(11,10)^{3}$&  \\ \hline

		$(11,9) \times (10,10)^{2}$ 
		&\multirow{3}{*}{  1,3,9,21,48,99}  \\ \cline{1-1}
		$(12,8) \times (11,9)^{2}$ & \\ \cline{1-1}
		$(12,9) \times (11,10)^{2}$ & \\ \hline
		
		$(10,10+ \textcolor{black}{q}) \times (10,10)^{2}$ &\multirow{3}{*}{ 1,3,9,22,51,108}  \\ \cline{1-1}
		$(11,9+ \textcolor{black}{q}) \times (11,9)^{2}$&  \\ \cline{1-1}
		$(11,10+ \textcolor{black}{q}) \times (11,10)^{2}$&   \\ \hline
		
		$(11,9) \times (10,10)\times (10,10+q)$ 
		&\multirow{3}{*}{  1,4,11,28,63,134}  \\ \cline{1-1}
		$(12,8) \times (11,9) \times (11,9+q)$ &   \\ \cline{1-1}
		$(12,9) \times (11,10)\times (11,10+q)$ &  \\ \hline
	\end{tabular}
	\caption{Different LL configurations for parton state $\phi_2^3$ given by  $(N^{1}_{\rm LLL},N^{1}_{\rm LL1})\times (N^{2}_{\rm LLL},N^{2}_{\rm LL1})\times (N^{3}_{\rm LLL},N^{3}_{\rm LL1})$ along with counting corresponding to the dominant spectra of its overlap matrix $\mathcal{O}$.  $(N^{i}_{\rm LLL},N^{i}_{\rm LL1}+q)$ represents the LL occupation sector where a quasihole is added to LL1 of the $i$th parton.}
	\label{tab2}
\end{table}

\section{Conclusion}
\label{sec: conclusion}

We have studied the RSES of a variety of unprojected parton states given by $\phi_2^2,\; \phi_2^3$ and $\phi_3^2$ for bipartitions of systems with as many as $80$ particles. The RSES of $\phi_2^2$ and $\phi_2^3$ was also computed when the parton states had quasihole excitations in the bulk. We found that counting present in RSES for the unprojected parton wave function of the form $\phi_n^k$ has a one-to-one mapping with the counting of states in the $\widehat{su}(n)_k \times u(1)$ edge current algebra. RSES counting that match with different representations of the algebra could be realized by considering sectors of the reduced density matrix with odd and even numbers of particles or by insertion of a quasihole.

We also studied RSES of LLL-projected $[\phi_1^2 \phi_2]^2$ state with an approximate projection scheme. We found that the RSES for this state has an identical structure as that of unprojected $\phi_2^2$ which suggests that both states are topologically equivalent (up to Abelian anyons). Multiplication with $\phi_1^2$ and LLL-projection does not change the topological class of a parton state (to be precise, multiplication by $\phi_1^2$ does not change the chiral central charge but does alter other topological quantities like the shift and the charges of the quasiparticles but these can be readily accounted for).

Finally, we computed the spectra of the overlap matrices corresponding to the edge excitations of the parton states while restricting the number of particles in parton Landau levels. The calculations indicate that the different branches seen in the RSES can be associated with different Landau level occupancies of the partons. However, a more thorough study is needed to consider a wider range of Landau-level occupancies to see how they are embedded in the RSES. A careful study of the overlaps between the fixed Landau level occupancy edge state space and the eigenspace of states formed by the individual branches in the RSES can provide a more concrete demonstration of the connection between the Landau level occupancies and the branches. The qualitative similarity in the branch structure of the RSES to the ones previously studied \cite{Greg2021} for the Jain sequence states motivates the question of whether these entanglement spectra as well can be associated with a local entanglement Hamiltonian.

\begin{acknowledgments}
We thank Greg Henderson, Sunil Mukhi, Naveen Prabhakar, Jainendra Jain, Steve Simon, and Joost Slingerland for useful inputs. The work benefited from discussions at the International Centre for Theoretical Sciences program - Novel phases of quantum matter (Code: ICTS/topmatter2019/12) and Geometric phases in Optics and Topological Matter (Code: ICTS/geomtop2020/1). A. A. is supported by SRF-CSIR (India), Grant No. 09/936(0220)/2019-EMR-I. A.C.B.  acknowledges the Science and Engineering Research Board (SERB) of the Department of Science and Technology (DST) for funding support via the Start-up Grant SRG/2020/000154. S.G.J. acknowledges DST-SERB (India) Grant No. ECR/2018/001781, and thanks LPT Toulouse for a visit funded by the grant NanoX ANR-17-EURE-0009 in the framework of the ``Programme des Investissements d’Avenir”, during which this manuscript was completed. The calculations were performed on the Param Brahma (NSM, IISER Pune) supercomputing facility.
\end{acknowledgments}

\bibliography{biblio_fqhe.bib}

\begin{thebibliography}{44}
\expandafter\ifx\csname natexlab\endcsname\relax\def\natexlab#1{#1}\fi
\expandafter\ifx\csname bibnamefont\endcsname\relax
  \def\bibnamefont#1{#1}\fi
\expandafter\ifx\csname bibfnamefont\endcsname\relax
  \def\bibfnamefont#1{#1}\fi
\expandafter\ifx\csname citenamefont\endcsname\relax
  \def\citenamefont#1{#1}\fi
\expandafter\ifx\csname url\endcsname\relax
  \def\url#1{\texttt{#1}}\fi
\expandafter\ifx\csname urlprefix\endcsname\relax\def\urlprefix{URL }\fi
\providecommand{\bibinfo}[2]{#2}
\providecommand{\eprint}[2][]{\url{#2}}

\bibitem[{\citenamefont{Jain}(1989{\natexlab{a}})}]{Jain89}
\bibinfo{author}{\bibfnamefont{J.~K.} \bibnamefont{Jain}},
  \bibinfo{journal}{Phys. Rev. Lett.} \textbf{\bibinfo{volume}{63}},
  \bibinfo{pages}{199} (\bibinfo{year}{1989}{\natexlab{a}}),
  \urlprefix\url{http://link.aps.org/doi/10.1103/PhysRevLett.63.199}.

\bibitem[{\citenamefont{Jain}(1989{\natexlab{b}})}]{Jain89b}
\bibinfo{author}{\bibfnamefont{J.~K.} \bibnamefont{Jain}},
  \bibinfo{journal}{Phys. Rev. B} \textbf{\bibinfo{volume}{40}},
  \bibinfo{pages}{8079} (\bibinfo{year}{1989}{\natexlab{b}}),
  \urlprefix\url{http://link.aps.org/doi/10.1103/PhysRevB.40.8079}.

\bibitem[{\citenamefont{Balram et~al.}(2018{\natexlab{a}})\citenamefont{Balram,
  Barkeshli, and Rudner}}]{Balram18}
\bibinfo{author}{\bibfnamefont{A.~C.} \bibnamefont{Balram}},
  \bibinfo{author}{\bibfnamefont{M.}~\bibnamefont{Barkeshli}},
  \bibnamefont{and} \bibinfo{author}{\bibfnamefont{M.~S.}
  \bibnamefont{Rudner}}, \bibinfo{journal}{Phys. Rev. B}
  \textbf{\bibinfo{volume}{98}}, \bibinfo{pages}{035127}
  (\bibinfo{year}{2018}{\natexlab{a}}),
  \urlprefix\url{https://link.aps.org/doi/10.1103/PhysRevB.98.035127}.

\bibitem[{\citenamefont{Balram et~al.}(2018{\natexlab{b}})\citenamefont{Balram,
  Mukherjee, Park, Barkeshli, Rudner, and Jain}}]{Balram18a}
\bibinfo{author}{\bibfnamefont{A.~C.} \bibnamefont{Balram}},
  \bibinfo{author}{\bibfnamefont{S.}~\bibnamefont{Mukherjee}},
  \bibinfo{author}{\bibfnamefont{K.}~\bibnamefont{Park}},
  \bibinfo{author}{\bibfnamefont{M.}~\bibnamefont{Barkeshli}},
  \bibinfo{author}{\bibfnamefont{M.~S.} \bibnamefont{Rudner}},
  \bibnamefont{and} \bibinfo{author}{\bibfnamefont{J.~K.} \bibnamefont{Jain}},
  \bibinfo{journal}{Phys. Rev. Lett.} \textbf{\bibinfo{volume}{121}},
  \bibinfo{pages}{186601} (\bibinfo{year}{2018}{\natexlab{b}}),
  \urlprefix\url{https://link.aps.org/doi/10.1103/PhysRevLett.121.186601}.

\bibitem[{\citenamefont{Balram et~al.}(2019)\citenamefont{Balram, Barkeshli,
  and Rudner}}]{Balram19}
\bibinfo{author}{\bibfnamefont{A.~C.} \bibnamefont{Balram}},
  \bibinfo{author}{\bibfnamefont{M.}~\bibnamefont{Barkeshli}},
  \bibnamefont{and} \bibinfo{author}{\bibfnamefont{M.~S.}
  \bibnamefont{Rudner}}, \bibinfo{journal}{Phys. Rev. B}
  \textbf{\bibinfo{volume}{99}}, \bibinfo{pages}{241108}
  (\bibinfo{year}{2019}),
  \urlprefix\url{https://link.aps.org/doi/10.1103/PhysRevB.99.241108}.

\bibitem[{\citenamefont{Balram et~al.}(2020)\citenamefont{Balram, Jain, and
  Barkeshli}}]{Balram19a}
\bibinfo{author}{\bibfnamefont{A.~C.} \bibnamefont{Balram}},
  \bibinfo{author}{\bibfnamefont{J.~K.} \bibnamefont{Jain}}, \bibnamefont{and}
  \bibinfo{author}{\bibfnamefont{M.}~\bibnamefont{Barkeshli}},
  \bibinfo{journal}{Phys. Rev. Research} \textbf{\bibinfo{volume}{2}},
  \bibinfo{pages}{013349} (\bibinfo{year}{2020}),
  \urlprefix\url{https://link.aps.org/doi/10.1103/PhysRevResearch.2.013349}.

\bibitem[{\citenamefont{Balram}(2021{\natexlab{a}})}]{Balram20a}
\bibinfo{author}{\bibfnamefont{A.~C.} \bibnamefont{Balram}},
  \bibinfo{journal}{SciPost Phys.} \textbf{\bibinfo{volume}{10}},
  \bibinfo{pages}{83} (\bibinfo{year}{2021}{\natexlab{a}}),
  \urlprefix\url{https://scipost.org/10.21468/SciPostPhys.10.4.083}.

\bibitem[{\citenamefont{Faugno et~al.}(2020)\citenamefont{Faugno, Jain, and
  Balram}}]{Faugno20a}
\bibinfo{author}{\bibfnamefont{W.~N.} \bibnamefont{Faugno}},
  \bibinfo{author}{\bibfnamefont{J.~K.} \bibnamefont{Jain}}, \bibnamefont{and}
  \bibinfo{author}{\bibfnamefont{A.~C.} \bibnamefont{Balram}},
  \bibinfo{journal}{Phys. Rev. Research} \textbf{\bibinfo{volume}{2}},
  \bibinfo{pages}{033223} (\bibinfo{year}{2020}),
  \urlprefix\url{https://link.aps.org/doi/10.1103/PhysRevResearch.2.033223}.

\bibitem[{\citenamefont{Faugno et~al.}(2021)\citenamefont{Faugno, Zhao, Balram,
  Jolicoeur, and Jain}}]{Faugno20b}
\bibinfo{author}{\bibfnamefont{W.~N.} \bibnamefont{Faugno}},
  \bibinfo{author}{\bibfnamefont{T.}~\bibnamefont{Zhao}},
  \bibinfo{author}{\bibfnamefont{A.~C.} \bibnamefont{Balram}},
  \bibinfo{author}{\bibfnamefont{T.}~\bibnamefont{Jolicoeur}},
  \bibnamefont{and} \bibinfo{author}{\bibfnamefont{J.~K.} \bibnamefont{Jain}},
  \bibinfo{journal}{Phys. Rev. B} \textbf{\bibinfo{volume}{103}},
  \bibinfo{pages}{085303} (\bibinfo{year}{2021}),
  \urlprefix\url{https://link.aps.org/doi/10.1103/PhysRevB.103.085303}.

\bibitem[{\citenamefont{Balram}(2022)}]{Balram21b}
\bibinfo{author}{\bibfnamefont{A.~C.} \bibnamefont{Balram}},
  \bibinfo{journal}{Phys. Rev. B} \textbf{\bibinfo{volume}{105}},
  \bibinfo{pages}{L121406} (\bibinfo{year}{2022}),
  \urlprefix\url{https://link.aps.org/doi/10.1103/PhysRevB.105.L121406}.

\bibitem[{\citenamefont{Wu et~al.}(2017)\citenamefont{Wu, Shi, and
  Jain}}]{Wu16}
\bibinfo{author}{\bibfnamefont{Y.}~\bibnamefont{Wu}},
  \bibinfo{author}{\bibfnamefont{T.}~\bibnamefont{Shi}}, \bibnamefont{and}
  \bibinfo{author}{\bibfnamefont{J.~K.} \bibnamefont{Jain}},
  \bibinfo{journal}{Nano Letters} \textbf{\bibinfo{volume}{17}},
  \bibinfo{pages}{4643} (\bibinfo{year}{2017}), \bibinfo{note}{pMID: 28649831},
  \eprint{http://dx.doi.org/10.1021/acs.nanolett.7b01080},
  \urlprefix\url{http://dx.doi.org/10.1021/acs.nanolett.7b01080}.

\bibitem[{\citenamefont{Kim et~al.}(2019)\citenamefont{Kim, Balram, Taniguchi,
  Watanabe, Jain, and Smet}}]{Kim18}
\bibinfo{author}{\bibfnamefont{Y.}~\bibnamefont{Kim}},
  \bibinfo{author}{\bibfnamefont{A.~C.} \bibnamefont{Balram}},
  \bibinfo{author}{\bibfnamefont{T.}~\bibnamefont{Taniguchi}},
  \bibinfo{author}{\bibfnamefont{K.}~\bibnamefont{Watanabe}},
  \bibinfo{author}{\bibfnamefont{J.~K.} \bibnamefont{Jain}}, \bibnamefont{and}
  \bibinfo{author}{\bibfnamefont{J.~H.} \bibnamefont{Smet}},
  \bibinfo{journal}{Nature Physics} \textbf{\bibinfo{volume}{15}},
  \bibinfo{pages}{154} (\bibinfo{year}{2019}), ISSN \bibinfo{issn}{1745-2481},
  \urlprefix\url{https://doi.org/10.1038/s41567-018-0355-x}.

\bibitem[{\citenamefont{Faugno et~al.}(2019)\citenamefont{Faugno, Balram,
  Barkeshli, and Jain}}]{Faugno19}
\bibinfo{author}{\bibfnamefont{W.~N.} \bibnamefont{Faugno}},
  \bibinfo{author}{\bibfnamefont{A.~C.} \bibnamefont{Balram}},
  \bibinfo{author}{\bibfnamefont{M.}~\bibnamefont{Barkeshli}},
  \bibnamefont{and} \bibinfo{author}{\bibfnamefont{J.~K.} \bibnamefont{Jain}},
  \bibinfo{journal}{Phys. Rev. Lett.} \textbf{\bibinfo{volume}{123}},
  \bibinfo{pages}{016802} (\bibinfo{year}{2019}),
  \urlprefix\url{https://link.aps.org/doi/10.1103/PhysRevLett.123.016802}.

\bibitem[{\citenamefont{Balram}(2021{\natexlab{b}})}]{Balram21}
\bibinfo{author}{\bibfnamefont{A.~C.} \bibnamefont{Balram}},
  \bibinfo{journal}{Phys. Rev. B} \textbf{\bibinfo{volume}{103}},
  \bibinfo{pages}{155103} (\bibinfo{year}{2021}{\natexlab{b}}),
  \urlprefix\url{https://link.aps.org/doi/10.1103/PhysRevB.103.155103}.

\bibitem[{\citenamefont{Balram and W\'ojs}(2021)}]{Balram21a}
\bibinfo{author}{\bibfnamefont{A.~C.} \bibnamefont{Balram}} \bibnamefont{and}
  \bibinfo{author}{\bibfnamefont{A.}~\bibnamefont{W\'ojs}},
  \bibinfo{journal}{Phys. Rev. Research} \textbf{\bibinfo{volume}{3}},
  \bibinfo{pages}{033087} (\bibinfo{year}{2021}),
  \urlprefix\url{https://link.aps.org/doi/10.1103/PhysRevResearch.3.033087}.

\bibitem[{\citenamefont{Balram et~al.}(2022)\citenamefont{Balram, Liu, Gromov,
  and Papi\ifmmode~\acute{c}\else \'{c}\fi{}}}]{Balram21d}
\bibinfo{author}{\bibfnamefont{A.~C.} \bibnamefont{Balram}},
  \bibinfo{author}{\bibfnamefont{Z.}~\bibnamefont{Liu}},
  \bibinfo{author}{\bibfnamefont{A.}~\bibnamefont{Gromov}}, \bibnamefont{and}
  \bibinfo{author}{\bibfnamefont{Z.}~\bibnamefont{Papi\ifmmode~\acute{c}\else
  \'{c}\fi{}}}, \bibinfo{journal}{Phys. Rev. X} \textbf{\bibinfo{volume}{12}},
  \bibinfo{pages}{021008} (\bibinfo{year}{2022}),
  \urlprefix\url{https://link.aps.org/doi/10.1103/PhysRevX.12.021008}.

\bibitem[{\citenamefont{Dora and Balram}(2022)}]{Dora22}
\bibinfo{author}{\bibfnamefont{R.~K.} \bibnamefont{Dora}} \bibnamefont{and}
  \bibinfo{author}{\bibfnamefont{A.~C.} \bibnamefont{Balram}},
  \emph{\bibinfo{title}{Nature of the anomalous $4/13$ fractional quantum
  {Hall} effect in graphene}} (\bibinfo{year}{2022}), \eprint{2202.10395}.

\bibitem[{\citenamefont{Haldane}(1983)}]{Haldane83}
\bibinfo{author}{\bibfnamefont{F.~D.~M.} \bibnamefont{Haldane}},
  \bibinfo{journal}{Phys. Rev. Lett.} \textbf{\bibinfo{volume}{51}},
  \bibinfo{pages}{605} (\bibinfo{year}{1983}),
  \urlprefix\url{http://link.aps.org/doi/10.1103/PhysRevLett.51.605}.

\bibitem[{\citenamefont{Wen and Zee}(1992)}]{Wen92}
\bibinfo{author}{\bibfnamefont{X.~G.} \bibnamefont{Wen}} \bibnamefont{and}
  \bibinfo{author}{\bibfnamefont{A.}~\bibnamefont{Zee}},
  \bibinfo{journal}{Phys. Rev. Lett.} \textbf{\bibinfo{volume}{69}},
  \bibinfo{pages}{953} (\bibinfo{year}{1992}),
  \urlprefix\url{http://link.aps.org/doi/10.1103/PhysRevLett.69.953}.

\bibitem[{\citenamefont{Wen}(1991)}]{Wen91}
\bibinfo{author}{\bibfnamefont{X.~G.} \bibnamefont{Wen}},
  \bibinfo{journal}{Phys. Rev. Lett.} \textbf{\bibinfo{volume}{66}},
  \bibinfo{pages}{802} (\bibinfo{year}{1991}),
  \urlprefix\url{http://link.aps.org/doi/10.1103/PhysRevLett.66.802}.

\bibitem[{\citenamefont{Laughlin}(1983)}]{Laughlin83}
\bibinfo{author}{\bibfnamefont{R.~B.} \bibnamefont{Laughlin}},
  \bibinfo{journal}{Phys. Rev. Lett.} \textbf{\bibinfo{volume}{50}},
  \bibinfo{pages}{1395} (\bibinfo{year}{1983}),
  \urlprefix\url{http://link.aps.org/doi/10.1103/PhysRevLett.50.1395}.

\bibitem[{\citenamefont{Wen}(1995)}]{Wen95}
\bibinfo{author}{\bibfnamefont{X.-G.} \bibnamefont{Wen}},
  \bibinfo{journal}{Advances in Physics} \textbf{\bibinfo{volume}{44}},
  \bibinfo{pages}{405} (\bibinfo{year}{1995}),
  \eprint{http://www.tandfonline.com/doi/pdf/10.1080/00018739500101566},
  \urlprefix\url{http://www.tandfonline.com/doi/abs/10.1080/00018739500101566}.

\bibitem[{\citenamefont{Li and Haldane}(2008)}]{Li08}
\bibinfo{author}{\bibfnamefont{H.}~\bibnamefont{Li}} \bibnamefont{and}
  \bibinfo{author}{\bibfnamefont{F.~D.~M.} \bibnamefont{Haldane}},
  \bibinfo{journal}{Phys. Rev. Lett.} \textbf{\bibinfo{volume}{101}},
  \bibinfo{pages}{010504} (\bibinfo{year}{2008}),
  \urlprefix\url{http://link.aps.org/doi/10.1103/PhysRevLett.101.010504}.

\bibitem[{\citenamefont{Francesco et~al.}(1997)\citenamefont{Francesco,
  Mathieu, and Senechal}}]{DiFrancesco97}
\bibinfo{author}{\bibfnamefont{P.~D.} \bibnamefont{Francesco}},
  \bibinfo{author}{\bibfnamefont{P.}~\bibnamefont{Mathieu}}, \bibnamefont{and}
  \bibinfo{author}{\bibfnamefont{D.}~\bibnamefont{Senechal}},
  \emph{\bibinfo{title}{Conformal Field Theory}}, Graduate Texts in
  Contemporary Physics (\bibinfo{publisher}{Springer}, \bibinfo{year}{1997}),
  ISBN \bibinfo{isbn}{9780387947853},
  \urlprefix\url{http://books.google.com/books?id=keUrdME5rhIC}.

\bibitem[{\citenamefont{Read and Rezayi}(1999)}]{Read99}
\bibinfo{author}{\bibfnamefont{N.}~\bibnamefont{Read}} \bibnamefont{and}
  \bibinfo{author}{\bibfnamefont{E.}~\bibnamefont{Rezayi}},
  \bibinfo{journal}{Phys. Rev. B} \textbf{\bibinfo{volume}{59}},
  \bibinfo{pages}{8084} (\bibinfo{year}{1999}),
  \urlprefix\url{http://link.aps.org/doi/10.1103/PhysRevB.59.8084}.

\bibitem[{\citenamefont{Zhu et~al.}(2015)\citenamefont{Zhu, Gong, Haldane, and
  Sheng}}]{Zhu15}
\bibinfo{author}{\bibfnamefont{W.}~\bibnamefont{Zhu}},
  \bibinfo{author}{\bibfnamefont{S.~S.} \bibnamefont{Gong}},
  \bibinfo{author}{\bibfnamefont{F.~D.~M.} \bibnamefont{Haldane}},
  \bibnamefont{and} \bibinfo{author}{\bibfnamefont{D.~N.} \bibnamefont{Sheng}},
  \bibinfo{journal}{Phys. Rev. Lett.} \textbf{\bibinfo{volume}{115}},
  \bibinfo{pages}{126805} (\bibinfo{year}{2015}),
  \urlprefix\url{http://link.aps.org/doi/10.1103/PhysRevLett.115.126805}.

\bibitem[{\citenamefont{Bandyopadhyay et~al.}(2018)\citenamefont{Bandyopadhyay,
  Chen, Ahari, Ortiz, Nussinov, and Seidel}}]{Bandyopadhyay18}
\bibinfo{author}{\bibfnamefont{S.}~\bibnamefont{Bandyopadhyay}},
  \bibinfo{author}{\bibfnamefont{L.}~\bibnamefont{Chen}},
  \bibinfo{author}{\bibfnamefont{M.~T.} \bibnamefont{Ahari}},
  \bibinfo{author}{\bibfnamefont{G.}~\bibnamefont{Ortiz}},
  \bibinfo{author}{\bibfnamefont{Z.}~\bibnamefont{Nussinov}}, \bibnamefont{and}
  \bibinfo{author}{\bibfnamefont{A.}~\bibnamefont{Seidel}},
  \bibinfo{journal}{Phys. Rev. B} \textbf{\bibinfo{volume}{98}},
  \bibinfo{pages}{161118} (\bibinfo{year}{2018}),
  \urlprefix\url{https://link.aps.org/doi/10.1103/PhysRevB.98.161118}.

\bibitem[{\citenamefont{Ahari et~al.}(2022)\citenamefont{Ahari, Bandyopadhyay,
  Nussinov, Seidel, and Ortiz}}]{Ahari22}
\bibinfo{author}{\bibfnamefont{M.~T.} \bibnamefont{Ahari}},
  \bibinfo{author}{\bibfnamefont{S.}~\bibnamefont{Bandyopadhyay}},
  \bibinfo{author}{\bibfnamefont{Z.}~\bibnamefont{Nussinov}},
  \bibinfo{author}{\bibfnamefont{A.}~\bibnamefont{Seidel}}, \bibnamefont{and}
  \bibinfo{author}{\bibfnamefont{G.}~\bibnamefont{Ortiz}},
  \emph{\bibinfo{title}{Partons as unique ground states of quantum hall parent
  hamiltonians: The case of fibonacci anyons}} (\bibinfo{year}{2022}),
  \urlprefix\url{https://arxiv.org/abs/2204.09684}.

\bibitem[{\citenamefont{Rodriguez et~al.}(2012)\citenamefont{Rodriguez,
  Sterdyniak, Hermanns, Slingerland, and Regnault}}]{Rodriguez12b}
\bibinfo{author}{\bibfnamefont{I.~D.} \bibnamefont{Rodriguez}},
  \bibinfo{author}{\bibfnamefont{A.}~\bibnamefont{Sterdyniak}},
  \bibinfo{author}{\bibfnamefont{M.}~\bibnamefont{Hermanns}},
  \bibinfo{author}{\bibfnamefont{J.~K.} \bibnamefont{Slingerland}},
  \bibnamefont{and} \bibinfo{author}{\bibfnamefont{N.}~\bibnamefont{Regnault}},
  \bibinfo{journal}{Phys. Rev. B} \textbf{\bibinfo{volume}{85}},
  \bibinfo{pages}{035128} (\bibinfo{year}{2012}),
  \urlprefix\url{https://link.aps.org/doi/10.1103/PhysRevB.85.035128}.

\bibitem[{\citenamefont{Chandran et~al.}(2011)\citenamefont{Chandran, Hermanns,
  Regnault, and Bernevig}}]{Bernevig2011}
\bibinfo{author}{\bibfnamefont{A.}~\bibnamefont{Chandran}},
  \bibinfo{author}{\bibfnamefont{M.}~\bibnamefont{Hermanns}},
  \bibinfo{author}{\bibfnamefont{N.}~\bibnamefont{Regnault}}, \bibnamefont{and}
  \bibinfo{author}{\bibfnamefont{B.~A.} \bibnamefont{Bernevig}},
  \bibinfo{journal}{Phys. Rev. B} \textbf{\bibinfo{volume}{84}},
  \bibinfo{pages}{205136} (\bibinfo{year}{2011}),
  \urlprefix\url{https://link.aps.org/doi/10.1103/PhysRevB.84.205136}.

\bibitem[{\citenamefont{Dubail et~al.}(2012)\citenamefont{Dubail, Read, and
  Rezayi}}]{DRR12}
\bibinfo{author}{\bibfnamefont{J.}~\bibnamefont{Dubail}},
  \bibinfo{author}{\bibfnamefont{N.}~\bibnamefont{Read}}, \bibnamefont{and}
  \bibinfo{author}{\bibfnamefont{E.~H.} \bibnamefont{Rezayi}},
  \bibinfo{journal}{Phys. Rev. B} \textbf{\bibinfo{volume}{85}},
  \bibinfo{pages}{115321} (\bibinfo{year}{2012}),
  \urlprefix\url{https://link.aps.org/doi/10.1103/PhysRevB.85.115321}.

\bibitem[{\citenamefont{Rodr\'{\i}guez and Sierra}(2009)}]{SierraRodriguez2009}
\bibinfo{author}{\bibfnamefont{I.~D.} \bibnamefont{Rodr\'{\i}guez}}
  \bibnamefont{and} \bibinfo{author}{\bibfnamefont{G.}~\bibnamefont{Sierra}},
  \bibinfo{journal}{Phys. Rev. B} \textbf{\bibinfo{volume}{80}},
  \bibinfo{pages}{153303} (\bibinfo{year}{2009}),
  \urlprefix\url{https://link.aps.org/doi/10.1103/PhysRevB.80.153303}.

\bibitem[{\citenamefont{Rodr\'{\i}guez
  et~al.}(2012)\citenamefont{Rodr\'{\i}guez, Simon, and
  Slingerland}}]{RSES:simon2012}
\bibinfo{author}{\bibfnamefont{I.~D.} \bibnamefont{Rodr\'{\i}guez}},
  \bibinfo{author}{\bibfnamefont{S.~H.} \bibnamefont{Simon}}, \bibnamefont{and}
  \bibinfo{author}{\bibfnamefont{J.~K.} \bibnamefont{Slingerland}},
  \bibinfo{journal}{Phys. Rev. Lett.} \textbf{\bibinfo{volume}{108}},
  \bibinfo{pages}{256806} (\bibinfo{year}{2012}),
  \urlprefix\url{https://link.aps.org/doi/10.1103/PhysRevLett.108.256806}.

\bibitem[{\citenamefont{Henderson
  et~al.}(2021{\natexlab{a}})\citenamefont{Henderson, Sreejith, and
  Simon}}]{RSES:2021}
\bibinfo{author}{\bibfnamefont{G.~J.} \bibnamefont{Henderson}},
  \bibinfo{author}{\bibfnamefont{G.~J.} \bibnamefont{Sreejith}},
  \bibnamefont{and} \bibinfo{author}{\bibfnamefont{S.~H.} \bibnamefont{Simon}},
  \bibinfo{journal}{Phys. Rev. B} \textbf{\bibinfo{volume}{104}},
  \bibinfo{pages}{195434} (\bibinfo{year}{2021}{\natexlab{a}}),
  \urlprefix\url{https://link.aps.org/doi/10.1103/PhysRevB.104.195434}.

\bibitem[{\citenamefont{Regnault}(2015)}]{Regnault2013}
\bibinfo{author}{\bibfnamefont{N.}~\bibnamefont{Regnault}},
  \emph{\bibinfo{title}{Entanglement spectroscopy and its application to the
  quantum hall effects}} (\bibinfo{year}{2015}),
  \urlprefix\url{https://arxiv.org/abs/1510.07670}.

\bibitem[{\citenamefont{Sreejith et~al.}(2018)\citenamefont{Sreejith, Fremling,
  Jeon, and Jain}}]{Sreejith18}
\bibinfo{author}{\bibfnamefont{G.~J.} \bibnamefont{Sreejith}},
  \bibinfo{author}{\bibfnamefont{M.}~\bibnamefont{Fremling}},
  \bibinfo{author}{\bibfnamefont{G.~S.} \bibnamefont{Jeon}}, \bibnamefont{and}
  \bibinfo{author}{\bibfnamefont{J.~K.} \bibnamefont{Jain}},
  \bibinfo{journal}{Phys. Rev. B} \textbf{\bibinfo{volume}{98}},
  \bibinfo{pages}{235139} (\bibinfo{year}{2018}),
  \urlprefix\url{https://link.aps.org/doi/10.1103/PhysRevB.98.235139}.

\bibitem[{\citenamefont{Henderson
  et~al.}(2021{\natexlab{b}})\citenamefont{Henderson, Sreejith, and
  Simon}}]{Greg2021}
\bibinfo{author}{\bibfnamefont{G.~J.} \bibnamefont{Henderson}},
  \bibinfo{author}{\bibfnamefont{G.~J.} \bibnamefont{Sreejith}},
  \bibnamefont{and} \bibinfo{author}{\bibfnamefont{S.~H.} \bibnamefont{Simon}},
  \bibinfo{journal}{Phys. Rev. B} \textbf{\bibinfo{volume}{104}},
  \bibinfo{pages}{195434} (\bibinfo{year}{2021}{\natexlab{b}}),
  \urlprefix\url{https://link.aps.org/doi/10.1103/PhysRevB.104.195434}.

\bibitem[{\citenamefont{Balram et~al.}(2015)\citenamefont{Balram, T\"oke,
  W\'ojs, and Jain}}]{Balram15a}
\bibinfo{author}{\bibfnamefont{A.~C.} \bibnamefont{Balram}},
  \bibinfo{author}{\bibfnamefont{C.}~\bibnamefont{T\"oke}},
  \bibinfo{author}{\bibfnamefont{A.}~\bibnamefont{W\'ojs}}, \bibnamefont{and}
  \bibinfo{author}{\bibfnamefont{J.~K.} \bibnamefont{Jain}},
  \bibinfo{journal}{Phys. Rev. B} \textbf{\bibinfo{volume}{92}},
  \bibinfo{pages}{075410} (\bibinfo{year}{2015}),
  \urlprefix\url{http://link.aps.org/doi/10.1103/PhysRevB.92.075410}.

\bibitem[{\citenamefont{Balram and Jain}(2016)}]{Balram16b}
\bibinfo{author}{\bibfnamefont{A.~C.} \bibnamefont{Balram}} \bibnamefont{and}
  \bibinfo{author}{\bibfnamefont{J.~K.} \bibnamefont{Jain}},
  \bibinfo{journal}{Phys. Rev. B} \textbf{\bibinfo{volume}{93}},
  \bibinfo{pages}{235152} (\bibinfo{year}{2016}),
  \urlprefix\url{http://link.aps.org/doi/10.1103/PhysRevB.93.235152}.

\bibitem[{\citenamefont{Girvin and Jach}(1984)}]{Girvin84b}
\bibinfo{author}{\bibfnamefont{S.~M.} \bibnamefont{Girvin}} \bibnamefont{and}
  \bibinfo{author}{\bibfnamefont{T.}~\bibnamefont{Jach}},
  \bibinfo{journal}{Phys. Rev. B} \textbf{\bibinfo{volume}{29}},
  \bibinfo{pages}{5617} (\bibinfo{year}{1984}),
  \urlprefix\url{http://link.aps.org/doi/10.1103/PhysRevB.29.5617}.

\bibitem[{\citenamefont{Jain}(2007)}]{Jain07}
\bibinfo{author}{\bibfnamefont{J.~K.} \bibnamefont{Jain}},
  \emph{\bibinfo{title}{Composite Fermions}} (\bibinfo{publisher}{Cambridge
  University Press, New York, US}, \bibinfo{year}{2007}).

\bibitem[{\citenamefont{Sreejith et~al.}(2011)\citenamefont{Sreejith, Jolad,
  Sen, and Jain}}]{Sreejith11c}
\bibinfo{author}{\bibfnamefont{G.~J.} \bibnamefont{Sreejith}},
  \bibinfo{author}{\bibfnamefont{S.}~\bibnamefont{Jolad}},
  \bibinfo{author}{\bibfnamefont{D.}~\bibnamefont{Sen}}, \bibnamefont{and}
  \bibinfo{author}{\bibfnamefont{J.~K.} \bibnamefont{Jain}},
  \bibinfo{journal}{Phys. Rev. B} \textbf{\bibinfo{volume}{84}},
  \bibinfo{pages}{245104} (\bibinfo{year}{2011}),
  \urlprefix\url{https://link.aps.org/doi/10.1103/PhysRevB.84.245104}.

\bibitem[{\citenamefont{Rodr\'{\i}guez
  et~al.}(2013)\citenamefont{Rodr\'{\i}guez, Davenport, Simon, and
  Slingerland}}]{Rodriguez13}
\bibinfo{author}{\bibfnamefont{I.~D.} \bibnamefont{Rodr\'{\i}guez}},
  \bibinfo{author}{\bibfnamefont{S.~C.} \bibnamefont{Davenport}},
  \bibinfo{author}{\bibfnamefont{S.~H.} \bibnamefont{Simon}}, \bibnamefont{and}
  \bibinfo{author}{\bibfnamefont{J.~K.} \bibnamefont{Slingerland}},
  \bibinfo{journal}{Phys. Rev. B} \textbf{\bibinfo{volume}{88}},
  \bibinfo{pages}{155307} (\bibinfo{year}{2013}),
  \urlprefix\url{https://link.aps.org/doi/10.1103/PhysRevB.88.155307}.

\bibitem[{\citenamefont{Kass et~al.}(1991)\citenamefont{Kass, Moody, Patera,
  and Slansky}}]{Kac1991}
\bibinfo{author}{\bibfnamefont{S.}~\bibnamefont{Kass}},
  \bibinfo{author}{\bibfnamefont{R.~V.} \bibnamefont{Moody}},
  \bibinfo{author}{\bibfnamefont{J.}~\bibnamefont{Patera}}, \bibnamefont{and}
  \bibinfo{author}{\bibfnamefont{R.}~\bibnamefont{Slansky}},
  \emph{\bibinfo{title}{Affine Lie Algebras, Weight Multiplicities, and
  Branching Rules (2 Volume Set)}} (\bibinfo{publisher}{University of
  California Press}, \bibinfo{year}{1991}), \bibinfo{edition}{hardcover} ed.,
  ISBN \bibinfo{isbn}{978-0520067684}.

\end{thebibliography}
\bibliographystyle{apsrev}

\pagebreak
\onecolumngrid

\begin{center}
	\textbf{\large Appendix}
\end{center}

\setcounter{equation}{0}
\setcounter{figure}{0}
\setcounter{table}{0}
\setcounter{page}{1}
\setcounter{section}{0}
\makeatletter
\renewcommand{\thesection}{A\arabic{section}}
\renewcommand{\theequation}{A\arabic{equation}}
\renewcommand{\thefigure}{A\arabic{figure}}
\renewcommand{\theHfigure}{A\arabic{figure}}
\renewcommand{\bibnumfmt}[1]{[A#1]}
\renewcommand{\citenumfont}[1]{#1}

\section{Representations of $\widehat{su}(2)_{2}$, $\widehat{su}(2)_{3}$ and $\widehat{su}(3)_{2}$ Kac-Moody algebra}
\label{app: rep_su2_level2}
  
\begin{figure}[h!]
	\centering
	\includegraphics[width=0.85\textwidth]{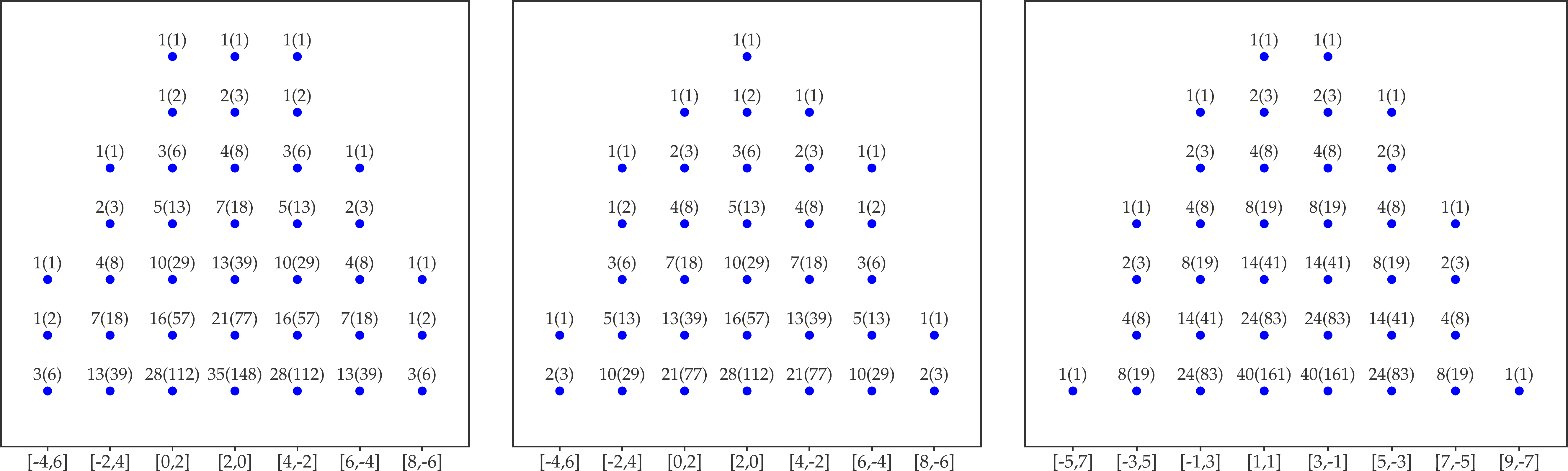}
	\caption{Representations of $\widehat{su}(2)_2$ for all three  highest weight states given by Dynkin label $[0,2]$ (left), $[2,0]$ (middle) and $[1,1]$. Dynkin labels for each state are labeled along $x$-axis. The number above each marker shows the multiplicity of the state whereas the number inside the parenthesis shows the multiplicity corresponding to $\widehat{su}(2)_2 \times {u}(1)$. 	}	
	\label{afig::su22All}
\end{figure}

A complete summary of root systems of affine extensions of ${su}(2)$ and ${su}(3)$ can be found in Ref.~\onlinecite{DiFrancesco97} (Also see explicit tables in Ref.~\onlinecite{Kac1991}). We used these ideas to construct the explicit multiplicities for the cases we are interested in. For ${su}(2)$ affine algebra, the simple roots in the Dynkin label basis are given by $E_{0}=[2,-2]$ and $E_{1}=[-2,2]$. In case of $\widehat{su}(2)_{2}$, the possible highest weights are $[2,0]$, $[1,1]$ and $[0,2]$. Following the procedure described, we get their highest weight representations (Fig.~\ref{afig::su22All}).

\begin{figure}[h]
	\includegraphics[width=0.85\textwidth]{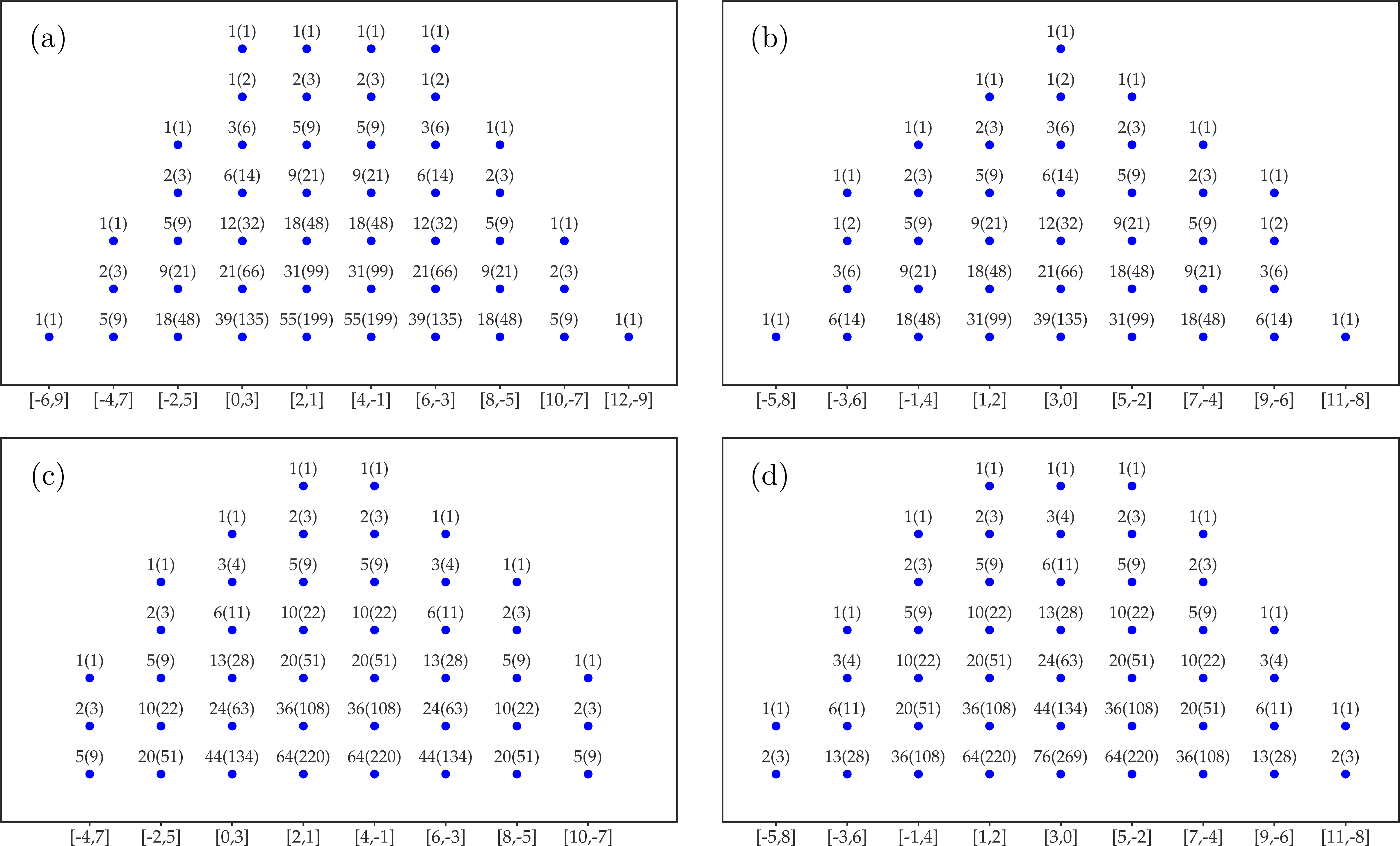}
	\caption{Representations of $\widehat{su}(2)_3$ for all possible highest weight states given by Dynkin label: (a) $[0,3]$, (b) $[3,0]$, (c) $[2,1]$ and (d) $[1,2]$. Dynkin labels for each state are labeled along $x$-axis. The number inside the parenthesis shows the multiplicity corresponding to $\widehat{su}(2)_3 \times {u}(1)$. 	}
	\label{afig::su23All}
\end{figure}

Similarly for $\widehat{su}(2)_{3}$, the possible highest weights are $[3,0]$, $[2,1]$, $[1,2]$ and $[0,3]$. The highest weight representations are presented in Fig.~\ref{afig::su23All}.

For ${su}(3)$ affine algebra, the simple roots are $E_{0}=[2,-1,-1]$, $E_{1}=[-1,2,-1]$ and $E_{2}=[-1,-1,2]$. For $\widehat{su}(3)_{2}$ the possible highest weights are $[2,0,0]$, $[0,2,0]$, $[0,0,2]$, $[1,1,0]$, $[1,0,1]$ and $[0,1,1]$. Following figures show the representation of $\widehat{su}(3)_{2}$ for different highest weight states:

\begin{figure}[h!]
	\includegraphics[width=0.9\textwidth]{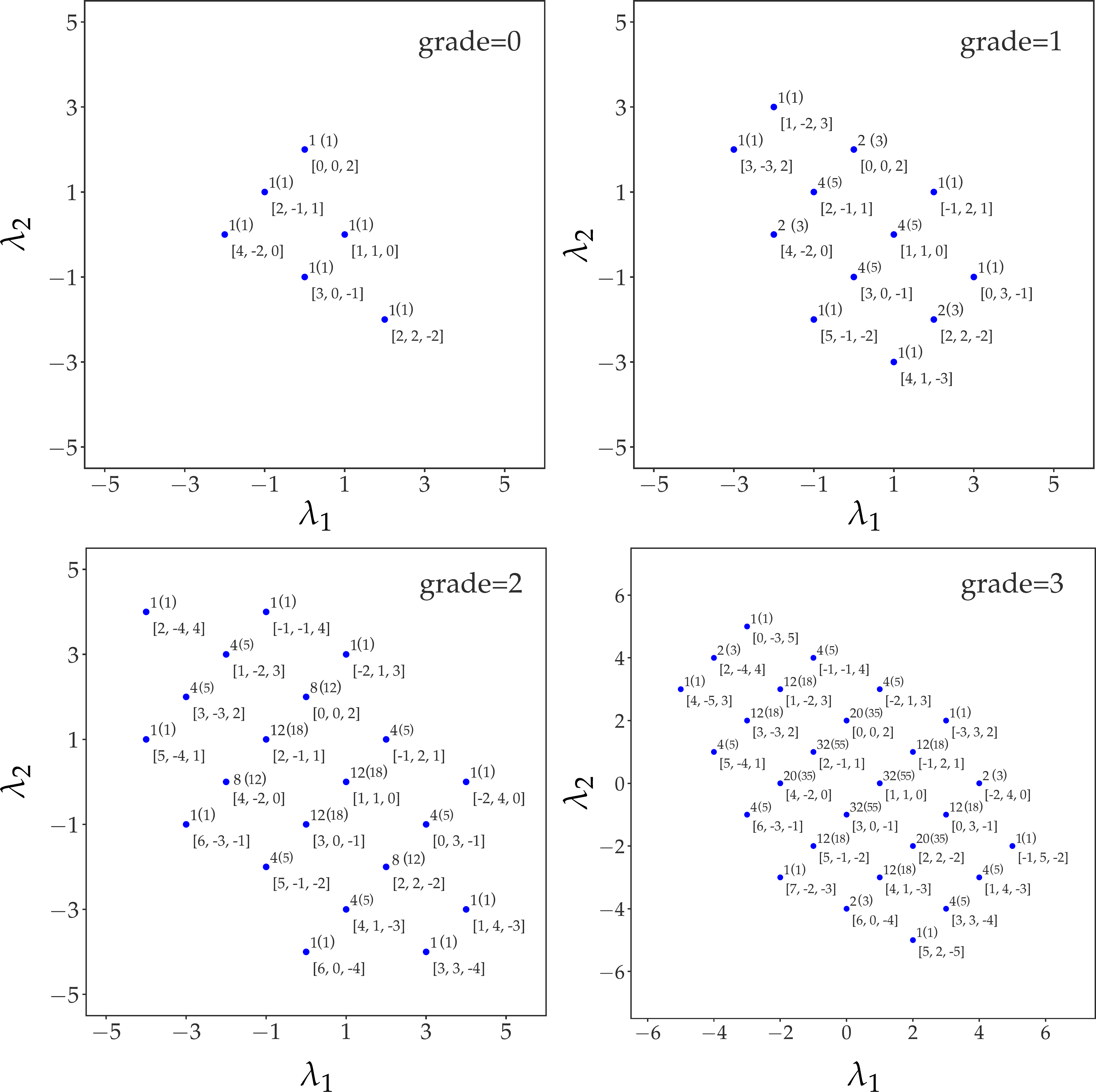}
	\caption{Representation of $\widehat{su}(3)_2$ for highest weight state given by Dynkin label $[0,0,2]$. This representation is identical to that for highest weight state given by Dynkin label $[0,2,0]$, but with a rotation. Each panel shows a particlular slice of the full representation labeled by grade quantum number.}
	\label{fig:su32_002}
\end{figure}

\begin{figure}[h!]
	\includegraphics[width=0.9\textwidth]{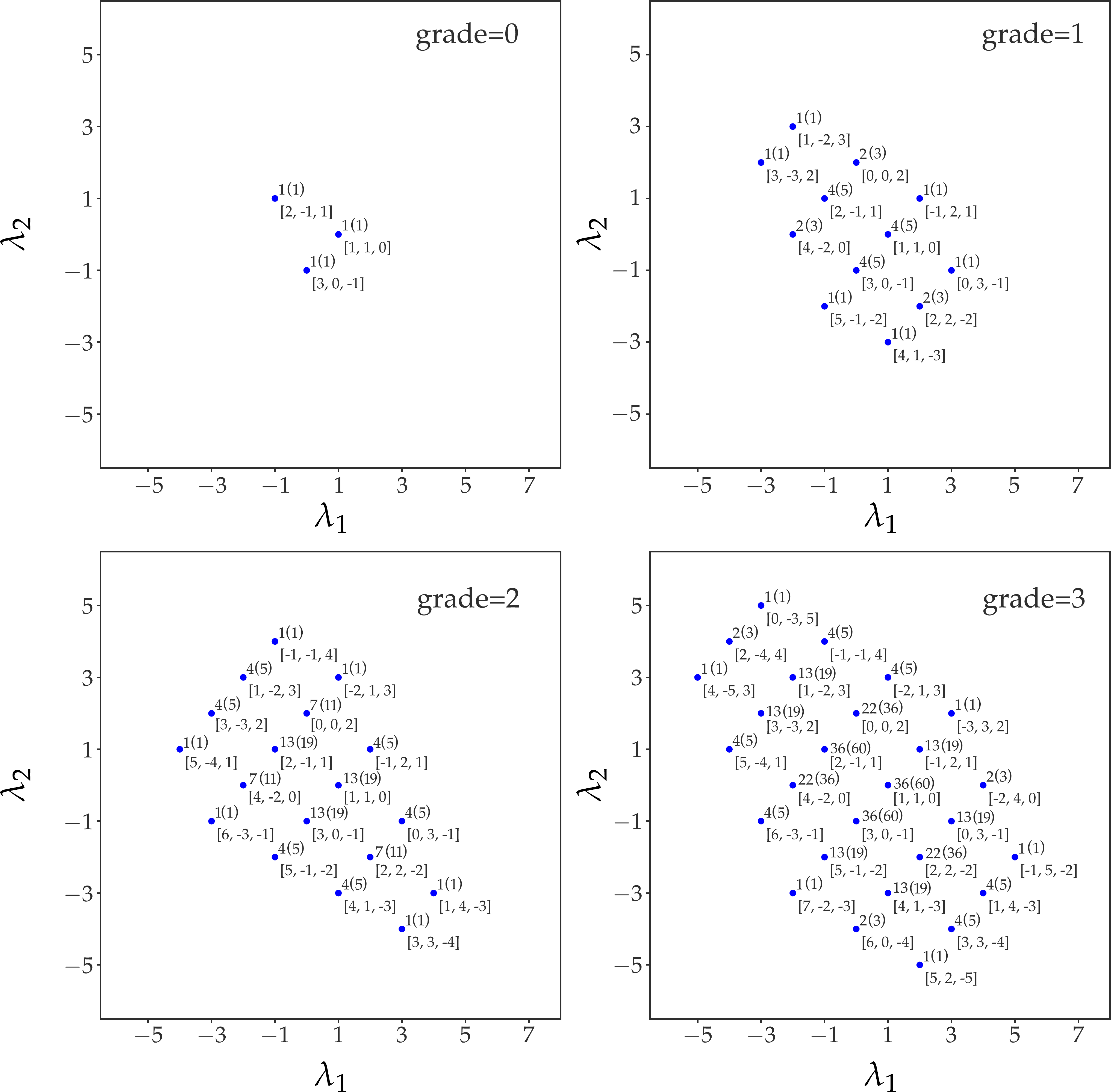}
	\caption{ Representation of $\widehat{su}(3)_2$ for highest weight state given by Dynkin label $[1,1,0]$. This representation is identical to that for highest weight state given by Dynkin label $[1,0,1]$, but with a rotation. Each panel shows a particlular slice of the full representation labeled by the grade quantum number.}
	\label{fig:su32_110}
\end{figure}

\begin{figure}[h!]
	\includegraphics[width=0.9\textwidth]{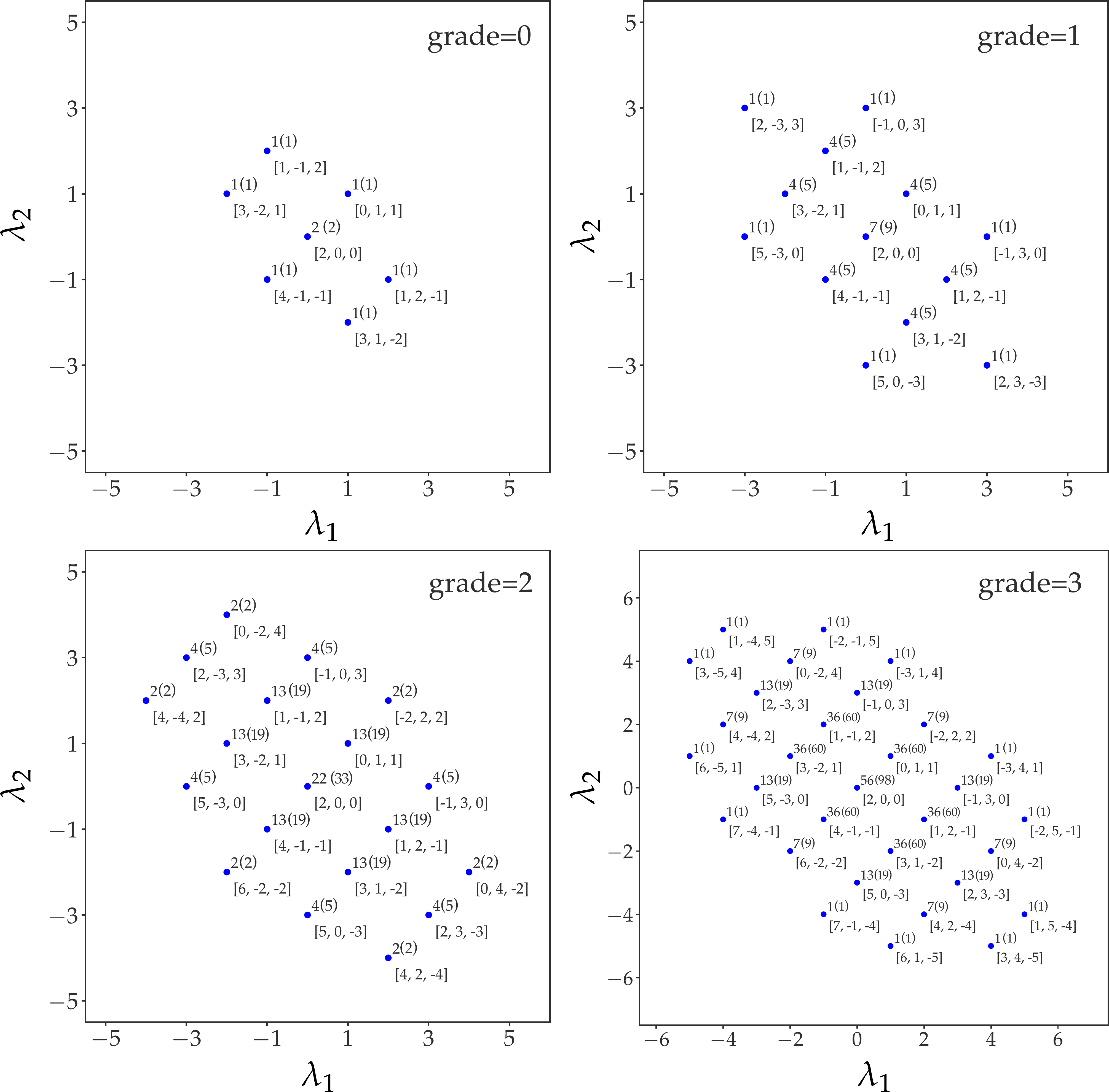}
	\caption{Representation of $\widehat{su}(3)_2$ for highest weight state given by Dynkin label $[0,1,1]$. Each panel shows a particlular slice of the full representation labeled by the grade quantum number. Action of $E_1$ and $E_2$ on $[0,1,1]$ state  creates large number of states in $\text{grade}=0$.}
	\label{fig:su32_011}
\end{figure}

\end{document}